\documentclass[sigconf]{acmart}

\AtBeginDocument{%
  }

\copyrightyear{}
\setcopyright{rightsretained}
\acmDOI{}
\acmISBN{}

\usepackage{acmart-taps}
\usepackage{listings}

\usepackage{isca25-125}
\usepackage{framed}
\usepackage{subcaption}
\usepackage{multirow}
\usepackage{makecell}
\usepackage{textcomp}
\usepackage[ruled,vlined,linesnumbered]{algorithm2e}
\usepackage{url}
\usepackage{pifont}

\newcommand{\highlight}[1]{\fontsize{10}{11}\texttt{#1}}
\newcommand{\Fig}[1]{Fig.~\ref{#1}}
\newcommand{\Tab}[1]{Table~\ref{#1}}
\newcommand{\Sec}[1]{\S\ref{#1}}
\newcommand{\eg}[0]{\textit{e.g.}\xspace}
\newcommand{\ie}[0]{\textit{i.e.}\xspace}
\newcommand{\aka}[0]{\textit{aka}\xspace}
\newcommand{\red}[1]{\textcolor{red}{#1}}
\newcommand{\blue}[1]{\textcolor{blue}{#1}}
\newcommand{\design}[0]{\textit{A4}\xspace}
\newcommand{\niparagraph}[1]{\noindent\textbf{#1}}

\begin{document}

\title{A4: Microarchitecture-Aware LLC Management for Datacenter Servers with Emerging I/O Devices}


\settopmatter{authorsperrow=3}




\author{Haneul Park}
\email{hnpark2@illinois.edu}
\affiliation{%
  \institution{University of Illinois, Urbana-Champaign}
  \city{Urbana}
  \state{IL}
  \country{United States}
}

\author{Jiaqi Lou}
\email{jiaqil6@illinois.edu}
\affiliation{%
  \institution{University of Illinois, Urbana-Champaign}
  \city{Urbana}
  \state{IL}
  \country{United States}
}

\author{Sangjin Lee}
\email{tkdwls0727@cau.ac.kr}
\affiliation{%
  \institution{Chung-Ang University}
  \city{Seoul}
  \state{}
  \country{Republic of Korea}
}

\author{Yifan Yuan}
\email{yifanyuan@meta.com}
\affiliation{%
  \institution{Meta}
  \city{Menlo Park}
  \state{CA}
  \country{United States}
}

\author{Kyoung Soo Park}
\email{kyoungsoo93@snu.ac.kr}
\affiliation{%
  \institution{Seoul National University}
  \city{Seoul}
  \state{}
  \country{Republic of Korea}
}

\author{Yongseok Son}
\email{sysganda@cau.ac.kr}
\affiliation{%
  \institution{Chung-Ang University}
  \city{Seoul}
  \state{}
  \country{Republic of Korea}
}

\author{Ipoom Jeong}
\email{ipoom@yonsei.ac.kr}
\affiliation{%
  \institution{Yonsei University}
  \city{Seoul}
  \state{}
  \country{Republic of Korea}
}

\author{Nam Sung Kim}
\email{nskim@illinois.edu}
\affiliation{%
  \institution{University of Illinois, Urbana-Champaign}
  \city{Urbana}
  \state{IL}
  \country{United States}
}

\renewcommand{\shortauthors}{Park, et al.}


\begin{abstract}

In modern server CPUs, the Last-Level Cache (LLC) serves not only as a victim cache for higher-level private caches but also as a buffer for low-latency DMA transfers between CPU cores and I/O devices through Direct Cache Access (DCA).
However, prior work has shown that high-bandwidth network-I/O devices can rapidly flood the LLC with packets, often causing significant contention with co-running workloads.
%
%
One step further, this work explores hidden microarchitectural properties of the Intel Xeon CPUs, uncovering two previously unrecognized LLC contentions triggered by emerging high-bandwidth I/O devices.
Specifically, (C1) DMA-written cache lines in LLC ways designated for DCA (referred to as DCA ways) are migrated to certain LLC ways (denoted as inclusive ways) when accessed by CPU cores, unexpectedly contending with non-I/O cache lines within the inclusive ways.
In addition, (C2) high-bandwidth storage-I/O devices, which are increasingly common in datacenter servers, benefit little from DCA while contending with (latency-sensitive) network-I/O devices within DCA ways. 
To this end, we present \design, a runtime LLC management framework designed to alleviate both (C1) and (C2) among diverse co-running workloads, using a hidden knob and other hardware features implemented in those CPUs.
Additionally, we demonstrate that \design can also alleviate other previously known network-I/O-driven LLC contentions.
Overall, it improves the performance of latency-sensitive, high-priority workloads by 51\% without notably compromising that of low-priority workloads. 

\end{abstract}




\maketitle
\section{Introduction}
\label{introduction}
The on-chip Last-Level Cache (LLC) in modern server CPUs plays an important role in providing high performance and energy efficiency for datacenter servers, with much lower latency and energy consumption per access than the off-chip memory. 
For decades, several innovations in LLC architecture, technology, and management have been introduced to further improve performance and energy efficiency~\cite{iyer2021advances}. 
The latest innovations include (1) the evolution of the LLC from inclusive to non-inclusive cache architectures~\cite{intel2019xeon:online}, (2) configurable LLC allocation per CPU core or process~\cite{GitHubin14:online}, and (3) Direct Cache Access (DCA) technology~\cite{huggahalli2005direct}.
They facilitate (1) more efficient utilization of limited LLC capacity, (2) less contention among workloads within the LLC, and (3) lower latency for DMA transfers between I/O devices and CPU cores directly through the LLC.
Such design trends are not limited to certain CPU vendors.
Major CPU vendors like Intel, AMD, and Arm have implemented DCA features, known as Data Direct I/O (DDIO)~\cite{IntelDa34:online}, Smart Data Cache Injection (SDCI)~\cite{zen5:sdci}, and cache stashing~\cite{ArmDynam59:online}, respectively.
Moreover, recent Intel Xeon CPUs and AMD Zen~4 CPUs have adopted the non-inclusive LLC~\cite{velten2022memory,yan2019attack,Zen4Micr69:online}.
While these solutions have proven effective, they have also introduced various unintended contentions within the LLC, especially for servers connected to high-bandwidth network-I/O devices that DMA-transfer packets to the LLC at hundreds of Gbps~\cite{tootoonchian2018resq,pismenny2023shring,farshin2020reexamining,yuan2021don,alian2022idio,vemmou2022patching}.
Prior work has identified the sources of these contentions within the LLC and eased them with software-only and hardware-assisted solutions. 
Yet, we observe that some contentions within the LLC remain unexplained and unresolved by previous work.
In this work, with the latest advances in understanding the architecture of recent server CPUs~\cite{yan2019attack} and improving the performance of storage-I/O devices~\cite{StorageP98:online} in mind, we uncover two sources of hidden \textbf{\underline{C}}ontentions within the LLC (\S\ref{sec:motivation}). 

\niparagraph{(C1) Contention between I/O and non-I/O cache lines within hidden inclusive ways.} 
In addition to the previously known network-I/O-driven contentions within LLC ways designated for DCA (referred to as `DCA ways' henceforth) (\Sec{subsec:ddio}), we uncover another contention between I/O and non-I/O cache lines within certain LLC ways (denoted as `inclusive ways' hereafter) (\Sec{subsec:directory_contention}). 
This contention arises from the unique directory architecture of the non-inclusive cache architecture (\Sec{subsec:non_inclusive_llc}). 
Contemporary Intel CPUs have two groups of directory ways to track cache-coherence states of cache lines in the LLC and/or the Mid-Level Caches (MLCs), with 11 and 12 ways dedicated to each group, respectively.
The two groups share two directory ways, each coupled one-to-one with inclusive ways.
These inclusive ways are the only LLC ways capable of holding cache lines simultaneously in both the LLC and MLCs~\cite{yan2019attack}.
We reveal that DMA-written, LLC-exclusive cache lines in DCA ways are migrated to the inclusive ways when accessed by CPU cores and brought into MLCs.
This causes a significant contention between I/O and non-I/O cache lines within inclusive ways, degrading the performance of both the I/O and non-I/O workloads when the non-I/O workloads are obliviously allocated to inclusive ways. 
We refer to such contention as `directory contention' henceforward.

\niparagraph{(C2) Contention between storage-I/O and network-I/O cache lines within DCA ways.} 
Although DCA was originally designed for network-I/O devices, it indiscriminately allows any I/O devices, including storage-I/O devices, to DMA-write I/O data to DCA ways.
In the past, when storage-I/O devices provided an order of magnitude lower bandwidth than network-I/O devices, contentions between storage-I/O and network-I/O cache lines within DCA ways were negligible.
However, as the bandwidth of storage-I/O devices has increased to the same order of magnitude as that of network-I/O devices (\eg, 116Gbps for an NVMe SSD~\cite{CrucialT705:online}), we hypothesize that storage-I/O cache lines can significantly interfere with network-I/O cache lines at DCA ways. 
To validate the hypothesis, we set up a server configured similarly to a class of datacenter servers with multiple high-bandwidth I/O devices and demonstrate the following (\S\ref{subsec:ddio_unfriendly_io}). 
DCA does not improve storage-I/O throughput, especially with large I/O blocks and deep I/O queues---typical strategies for maximizing storage-I/O throughput~\cite{lee2021asynchronous,bhattacharya2003asynchronous,tseng2021demystifying}.
Meanwhile, storage-I/O cache lines frequently evict network-I/O cache lines from DCA ways before those cache lines are consumed, which degrades the performance of network-I/O workloads.

Next, we propose \design, a runtime micro\underline{\textbf{a}}rchitecture-\underline{\textbf{a}}ware LLC m\underline{\textbf{a}}nagement fr\underline{\textbf{a}}mework, which helps users address both (C1) and (C2) when co-running workloads with varying priorities: High-Priority Workloads (HPWs) and Low-Priority Workloads (LPWs).
Specifically, \design provides two key \textbf{\underline{F}}unctions built with a little-known knob of DCA, Cache Allocation Technology (CAT)~\cite{GitHubin14:online}, and performance counters implemented in server CPUs (\S\ref{sec:smartllc_overview}).

\niparagraph{(F1) Preventing LPWs from being obliviously allocated to inclusive ways.}
To avoid (C1), \design starts by strategically eschewing the allocation of inclusive ways to LPWs. 
It then adaptively adjusts the number of standard ways (\ie, LLC ways excluding DCA and inclusive ways) allocated to these LPWs to keep performance metrics---such as LLC hit rates of HPWs---within ranges set to maximize the overall performance. 
Meanwhile, \design does not explicitly allocate HPWs to specific LLC ways.
Instead, it allows HPWs, whose performance is highly sensitive to accessible LLC capacity, to use as much LLC space as possible, including any unused capacity from the LLC ways allocated to LPWs.
This allocation strategy enables priority-based LLC partitioning between HPWs and LPWs while promoting efficient LLC sharing among workloads with the same priority. 
It remains effective even when the number of co-running processes exceeds the available number of LLC ways---a common challenge in prior work, improving the performance of HPWs by 30\% without notably degrading that of LPWs. 

\niparagraph{(F2) Selectively disabling DCA for storage-I/O devices.}
When detecting (C2), \design first exploits a little-known knob of Intel CPUs that can selectively disable DCA for storage-I/O devices. 
This will make \textit{antagonistic} storage-I/O LPWs get storage-I/O blocks through the device-memory-MLC path instead of the device-DCA-MLC path, eschewing a contention between network-I/O and storage-I/O cache lines in DCA ways.
After these storage-I/O cache lines are consumed, nonetheless, they will be eventually evicted to the LLC and contend with other cache lines in standard ways, a phenomenon known as DMA-bloat~\cite{alian2022idio}. 
To ease such a contention, it also adaptively decreases the number of standard ways allocated to these storage-I/O LPWs, named trash ways, to as few as one.
Overall, these strategies improve the performance of especially network-I/O HPWs by 63\%, without compromising the performance of storage-I/O LPWs that are insensitive to DCA and LLC capacity. 

Lastly, \design can be easily extended to mitigate other previously known network-I/O-driven contentions within the LLC (\S\ref{subsec:ddio}).
Specifically, it can address the latent contention within DCA ways~\cite{yuan2021don} by extending (F1) to stop non-I/O HPWs from using DCA ways.
This is achieved by extending (F1) to explicitly allocate non-I/O HPWs to all LLC ways but DCA ways, while I/O HPWs remain not explicitly assigned to any LLC ways.
%
\design can also ease the contention caused by DMA bloat from network-I/O workload by allocating network-I/O workloads only to trash ways in (F2) after detecting DMA bloat.
This stops DMA-bloated network-I/O cache lines from contending with non-I/O cache lines within other LLC ways.
%
%
Overall, \design improves the performance of HPWs by 51\% without compromising that of LPWs. 

\section{Background}
In this section, we briefly describe modern cache architecture, Direct Cache Access (DCA), and storage-I/O stack.

\subsection{Modern Cache Architecture}
\label{subsec:non_inclusive_llc}

\niparagraph{Non-inclusive LLC.}
With more cores on a CPU, the limited capacity of the traditional inclusive LLC becomes a performance bottleneck, especially in multi-tenant environments~\cite {iyer2021advances}.
Meanwhile, to provide both high performance and strong isolation among CPU cores (\ie, tenants), the capacity of private MLCs has increased, but that of the shared LLC cannot be proportionally increased under the chip-size constraint. 
As a solution, non-inclusive cache architectures have been adopted for LLCs, where CPU cores bring cache lines directly from memory to their private MLCs first upon LLC misses.
Later, they might evict these cache lines to the LLC, which serves as a victim cache. 
When accessing these cache lines again, the CPU cores not only bring them into the MLCs but also leave them in the LLC until they are evicted from the LLC. 
As such, it can utilize a given LLC capacity more flexibly and efficiently for frequently accessed (shared) cache lines. 

\begin{figure}[!t]
    \centering
    \includegraphics[width=\columnwidth]{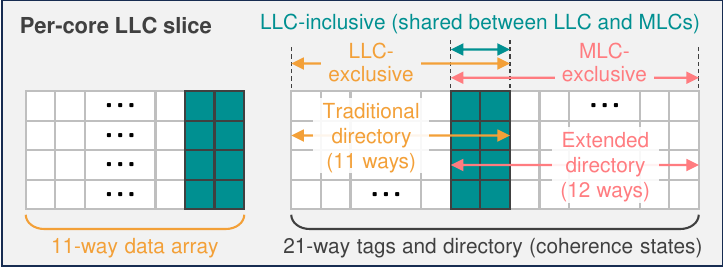}
    \caption{Reverse-engineered data array and directory structures of Intel Skylake CPUs~\cite{yan2019attack,wang2022understanding}.}
    \label{fig:skx_directory}
    \Description{Block diagram of an Intel Skylake LLC slice with 11‑way data array, traditional and extended directory groups. Two right‑most inclusive ways are highlighted as the only slots that can store a line present in both LLC and private caches.}
    \vspace{-10pt}
\end{figure}

\niparagraph{Inclusive directory.} 
Recent work has comprehensively reverse-engineered the directory architecture for the non-inclusive LLC of the Intel Skylake CPU and uncovered the following~\cite{yan2019attack}.
The directory architecture consists of (1) 11 traditional directory ways that are one-to-one coupled with 11 LLC ways and (2) 12 extended directory ways that track the cache coherence states of cache lines in MLCs (\Fig{fig:skx_directory}).
Two of the traditional directory ways are also used as two of the extended directory ways (the green region in \Fig{fig:skx_directory}). 
This gives rise to a coupling between the two hidden directory ways and two right-most LLC ways (also referred to as inclusive ways in this work).
LLC-inclusive cache lines (\ie, cache lines stored in both LLC and MLCs) are restricted to reside only in these two inclusive ways since other LLC ways cannot snoop MLCs, while LLC-exclusive cache lines may reside in any 11 ways.

\subsection{Direct Cache Access}
\label{subsec:ddio}

\begin{figure}[t]
    \centering
    \vspace{-0.5pt}
    \includegraphics[width=0.98\columnwidth]{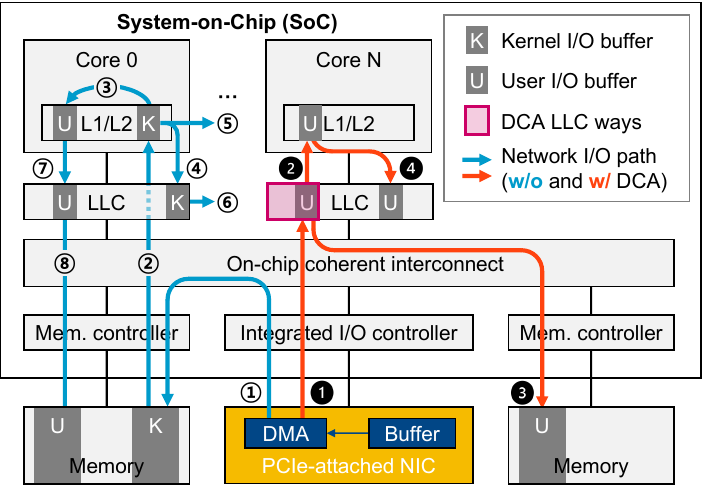}
        \caption{DMA paths of network I/O (blue: buffered I/O without DCA, red: kernel-bypass I/O with DCA)} 
    \label{fig:dma_paths_io}
    \Description{DMA data‑flow schematic: blue arrows show conventional device to memory and MLC path, red arrows show DCA device to LLC path.}
    \vspace{-10pt}
\end{figure}

The key insight behind the introduction of DCA is that the LLC, instead of the memory, can serve as the source and destination of the DMA transfer of I/O data between CPU cores and I/O devices. 
As CPU cores are likely to consume DMA-written I/O data soon, putting them close to the CPU improves latency and reduces bandwidth consumption.
If cache lines DMA-written by I/O devices are already present in the LLC, DCA performs in-place updates (\ie, write update). 
Otherwise, new cache lines are allocated to DCA ways---typically the two leftmost ways---and I/O data are DMA-written to these cache lines (\ie, write allocate).
While DCA was originally introduced for low-latency network-I/O devices, it also allows any (PCIe-based) I/O devices, such as NVMe SSDs~\cite{farshin2020reexamining,alian2020data,jeong2023ladio}, to DMA-transfer I/O data to/from DCA ways. 

\niparagraph{Ingress path.} 
The blue arrows in \Fig{fig:dma_paths_io} depict the ingress path of conventional I/O with a non-inclusive cache architecture.
After the allocation of kernel and user buffers, a given I/O device DMA-writes I/O data to the kernel buffer in the memory (\raisebox{.4pt}{\textcircled{\raisebox{-.7pt} {1}}}). 
Upon receiving an interrupt from the I/O device, CPU cores read the I/O data from the memory and then writes them to cache lines in their MLCs (\raisebox{.4pt}{\textcircled{\raisebox{-.7pt} {2}}});
we refer to cache lines storing I/O data as I/O cache lines henceforth.
Then the CPU cores copy these I/O cache lines to cache lines storing a user buffer (user-buffer cache lines) for processing (\raisebox{.4pt}{\textcircled{\raisebox{-.7pt} {3}}}).
Later, these I/O cache lines may be evicted to the LLC (\raisebox{.4pt}{\textcircled{\raisebox{-.7pt} {4}}}) or invalidated if reused/updated by the I/O device (\raisebox{.4pt}{\textcircled{\raisebox{-.7pt} {6}}}).
Similarly, the user-buffer cache lines may reside in the MLCs or be evicted to the LLC (\raisebox{.4pt}{\textcircled{\raisebox{-.9pt} {7}}}) and/or to the memory (\raisebox{.4pt}{\textcircled{\raisebox{-.9pt} {8}}}).
The red arrows show the ingress path of DCA-enabled, kernel-bypass network I/O such as DPDK~\cite{HomeDPDK50:online}, where a network-I/O device directly DMA-writes packets to cache lines in DCA ways (\raisebox{-0.3ex}{\scalebox{1.2}{\ding{108}}}\hspace{-1.7ex}\raisebox{0ex}{\textcolor{white}{1}}\hspace{0.6ex}) instead of the memory. 
Then, the CPU cores read these I/O cache lines from the DCA ways (\raisebox{-0.4ex}{\scalebox{1.2}{\ding{108}}}\hspace{-1.7ex}\raisebox{0ex}{\textcolor{white}{2}}\hspace{0.6ex}), obviating DMA-write to and then CPU-read from the memory (\raisebox{.4pt}{\textcircled{\raisebox{-.9pt} {1}}} and \raisebox{.4pt}{\textcircled{\raisebox{-.9pt} {2}}}, respectively).
Consequently, DCA effectively reduces the latency and bandwidth of accessing the memory for high-speed I/O. 
As these I/O cache lines 
are in a modified state and then written back to the memory (\raisebox{-0.3ex}{\scalebox{1.2}{\ding{108}}}\hspace{-1.6ex}\raisebox{0ex}{\textcolor{white}{3}}\hspace{0.6ex}), the coherence is maintained. 

\niparagraph{Egress path.}
The I/O data flow of the egress path is the opposite of that of the ingress path.
When I/O data written by CPU cores are only in their MLCs, I/O cache lines are copied to newly read-allocated cache lines in inclusive ways, and then DMA-read by I/O devices; if they are already in the LLC, they are DMA-read directly from the LLC~\cite{wang2022understanding}.
If I/O data are not in the cache hierarchy, I/O data are DMA-read by I/O devices directly from the memory, which does not read-allocate any cache lines in the LLC~\cite{kurth2020netcat}.
Since this behavior is not described in the architecture reference document, it is easily misinterpreted and leads to incorrect conclusions~\cite{purnal2022double}.

\niparagraph{Network-I/O-driven LLC contentions.} 
A body of work has demonstrated network-I/O-driven contentions within DCA ways and other LLC ways. 
The contention within DCA ways can be further divided into two contentions.
The first one, known as latent contention~\cite{yuan2021don}, occurs when a CPU core running a non-I/O workload is allocated to LLC ways that overlap with DCA ways, where cache lines storing network-I/O data (or simply network-I/O cache lines) contend with cache lines storing non-I/O data (or simply non-I/O cache lines) in DCA ways. 
The second one, referred to as DMA leak~\cite{farshin2020reexamining}, happens when network-I/O cache lines in DCA ways are evicted (\raisebox{-0.3ex}{\scalebox{1.2}{\ding{108}}}\hspace{-1.6ex}\raisebox{0ex}{\textcolor{white}{3}}\hspace{0.6ex}) by other network-I/O cache lines before they are consumed by CPU cores (\raisebox{-0.4ex}{\scalebox{1.2}{\ding{108}}}\hspace{-1.7ex}\raisebox{0ex}{\textcolor{white}{2}}\hspace{0.6ex}). 
The contention within other LLC ways can be caused by DMA bloat~\cite{alian2022idio}, occurring when consumed network-I/O cache lines in MLCs are evicted to LLC ways (\raisebox{-0.3ex}{\scalebox{1.2}{\ding{108}}}\hspace{-1.7ex}\raisebox{0ex}{\textcolor{white}{4}}\hspace{0.6ex}) where (DMA-bloated) I/O cache lines and non-I/O cache lines contend. 
This breaks the isolation that DCA aims to enforce, \ie, I/O cache lines remain in DCA ways when cached in the LLC.

\subsection{Modern Storage-I/O Stack}
\label{subsec:modern_file_io}
The Linux kernel and user-space NVMe drivers (\eg, SPDK~\cite{StorageP98:online}) have implemented software optimizations, such as Direct I/O~\cite{qian2024combining} and the use of large I/O blocks~\cite{lee2021asynchronous} with deeper I/O queues~\cite{bhattacharya2003asynchronous,tseng2021demystifying}, for the storage-I/O stack to maximize throughput while minimizing performance cost.
For example, Direct I/O is adopted by the standard Linux kernel and shares many features with Intel SPDK (\eg, kernel bypassing).
Similar to kernel-bypass network I/O, the Direct I/O (\eg, the \texttt{read()} system call with the \texttt{O\_DIRECT} flag) facilitates a workload to communicate with the SSD through a user buffer, bypassing the kernel page cache~\cite{park2017new}.
A body of work has investigated the interaction between network I/O and DCA~\cite{farshin2020reexamining,wang2022understanding,tootoonchian2018resq,pismenny2023shring,yuan2021don,alian2022idio,vemmou2022patching}.
Yet, less attention has been paid to the interaction between storage I/O and DCA, even though the NVMe SSD has begun to offer as high bandwidth as the NIC~\cite{DataCent58:online}. 
If the NIC in \Fig{fig:dma_paths_io} is replaced with storage, the DMA path of Direct I/O with DCA is analogous to that of the network I/O (red arrows).
This example assumes the user buffer has already been cached in LLC ways, and thus the corresponding I/O cache lines are write-updated in place (\raisebox{-0.3ex}{\scalebox{1.2}{\ding{108}}}\hspace{-1.65ex}\raisebox{0ex}{\textcolor{white}{1}}\hspace{0.6ex}).
Later, we show that 
storage I/O may cause significant contention between co-running workloads within both DCA ways and other LLC ways (\Sec{subsec:ddio_unfriendly_io}).
\section{Newly-discovered I/O-driven LLC Contentions}
\label{sec:motivation}
In this section, we first uncover a previously unrecognized I/O-driven contention within inclusive ways coupled with two directory ways in a non-inclusive cache architecture. 
Second, we identify an emerging contention between network and storage I/O within DCA ways, driven by the increasing storage-I/O bandwidth.

\subsection{Hidden Directory Contention}
\label{subsec:directory_contention}
\niparagraph{Setup.} 
We use the following two DPDK-based microbenchmarks as network-I/O workloads: (1) DPDK-\red{T} and (2) DPDK-\blue{NT}.
DPDK-T \textbf{\red{T}}ouches and then drops received network packets (\eg, deep packet inspection~\cite{LUCADERI84:online}). 
DPDK-NT does \textbf{\blue{N}}ot \textbf{\blue{T}}ouch network packets and simply drops them (\eg, packet classification and access control~\cite{59Packet3:online}). 
We run a given microbenchmark (\eg, DPDK-T) on four CPU cores, each with a 2K-entry ring buffer (\eg, totaling 8MB for storing 1KB packets), after allocating specific two consecutive LLC ways (\ie, way[5:6]) to the microbenchmark.
As a non-I/O workload, we run cache-sensitive X-Mem~\cite{GitHubmi87:online} with a 4MB working set on two CPU cores after allocating two consecutive LLC ways to X-Mem. 
The working set of X-Mem is larger than the aggregate capacity of two MLCs coupled with the two CPU cores but smaller than that of two LLC ways that we will allocate to X-Mem.
See \Sec{sec:methodology} for our system setup and control/monitor methods.

\begin{figure}[t]
    \centering
	\subfloat[DPDK-NT (w/o touching I/O data) vs. X-Mem\label{subfig:dpdk_ntch_xmem}]{%
        \includegraphics[width=\columnwidth]{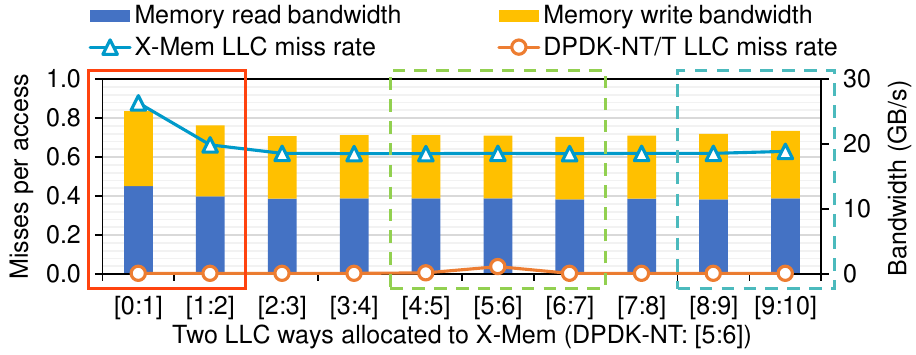}%
    }
    \hfill
    \vspace{6pt}
	\subfloat[DPDK-T (w/ touching I/O data) vs. X-Mem\label{subfig:dpdk_tch_xmem}]{%
    \includegraphics[width=\columnwidth]{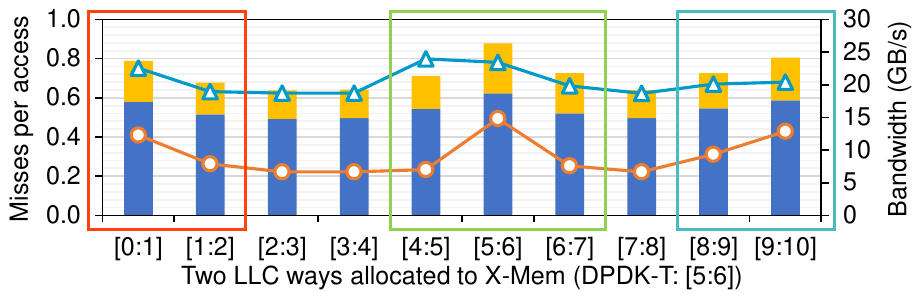}%
    }
    \Description{Sliding‑window plot of X‑Mem LLC miss rate and memory bandwidth while X-Mem's two‑way allocation moves across 11 LLC ways when co‑running DPDK‑NT. The Highest misses occur when the window overlaps DCA ways.}
    \vspace{-6pt}
    \caption{Contention between I/O-intensive DPDK and cache-sensitive X-Mem allocated to LLC way[\textit{m}:\textit{n}].}
    \Description{Same sliding‑window plot with DPDK‑T. Three distinct miss‑rate spikes at DCA ways, shared ways, and inclusive ways reveal latent, DMA‑bloat, and directory contention respectively.}
    \label{fig:ioa_ma_interference}
    \vspace{-10pt}
\end{figure}

\niparagraph{Contentions between I/O and non-I/O workloads.} 
\Fig{subfig:dpdk_ntch_xmem} shows the latent contention between DPDK-NT and X-Mem within DCA ways, where we sweep the allocation of two consecutive LLC ways to X-Mem from the two leftmost LLC ways (\ie, way[0:1] or DCA ways) to the two rightmost LLC ways (\ie, way[9:10] or inclusive ways), using Intel CAT~\cite{GitHubin14:online}.
This shows that X-Mem suffers higher LLC miss rates (also evidenced by higher consumption of memory bandwidth), when allocated to way[0:1] and way[1:2], which are fully and partially overlapped with DCA ways (red box in \Fig{subfig:dpdk_ntch_xmem}), respectively. 
This contention is caused by packets DMA-written to cache lines (referred to as network-I/O cache lines hereafter) that evict the cache lines of X-Mem in DCA ways. 
Meanwhile, X-mem does not experience notably higher LLC miss rates when allocated to way[5:6] which DPDK-NT is also allocated to.
This is because DPDK-NT does not bring I/O cache lines from DCA ways into its four MLCs since it does not consume them, preventing its working set from bloating beyond the capacity of four MLCs;
in other word, DPDK-NT does not evict I/O cache lines to way[5:6].

In comparison, \Fig{subfig:dpdk_tch_xmem} shows that X-Mem suffers higher LLC miss rates at three distinctive groups of two consecutive LLC ways allocated to X-Mem.   
The first group (red box in \Fig{subfig:dpdk_tch_xmem}) is caused by latent contention between DPDK-T and X-Mem within DCA ways, following the same contention mechanism as DPDK-NT (red box in \Fig{subfig:dpdk_ntch_xmem}).
The second group is incurred by contention between X-Mem and DPDK-T, both of which are (fully or partially) allocated to way[5:6] (green box in \Fig{subfig:dpdk_tch_xmem}). 
The source of this contention is DMA bloat caused by DPDK-T bringing fresh I/O cache lines from DCA ways into its four MLCs and subsequently evicting stale (or consumed) I/O cache lines to way[5:6].
This contrasts with DPDK-NT, which does not cause DMA-bloat to way[5:6] (dotted green box in \Fig{subfig:dpdk_ntch_xmem}).
Meanwhile, the third group (blue box in \Fig{subfig:dpdk_tch_xmem}) cannot be explained by the previously known contentions within the LLC.
Since the contention depicted by the third group is not observed when DPDK-NT co-runs with X-Mem (dotted blue box in \Fig{subfig:dpdk_ntch_xmem}), it must be caused by I/O cache lines brought into MLCs and then consumed by DPDK-T.  
Nonetheless, these cache lines should have been evicted only to way[5:6] allocated to DPDK-T, not to way[9:10]. 
We delve into this unexpected contention, inspired by the recent discovery of a microarchitectural property of a non-inclusive cache architecture with an inclusive directory (\Sec{subsec:non_inclusive_llc}).
Based on this, we hypothesize that LLC-exclusive cache lines in DCA ways are migrated to inclusive ways once they are read into MLCs.
This is supported by past work~\cite{wang2022understanding} showing that the cache coherence state of DMA-written cache lines changes from \emph{modified} LLC-exclusive to \emph{shared} LLC-inclusive when these cache lines are brought into MLCs and consumed by CPU cores.

\begin{figure}[t]
     \centering
     \includegraphics[width=\columnwidth]{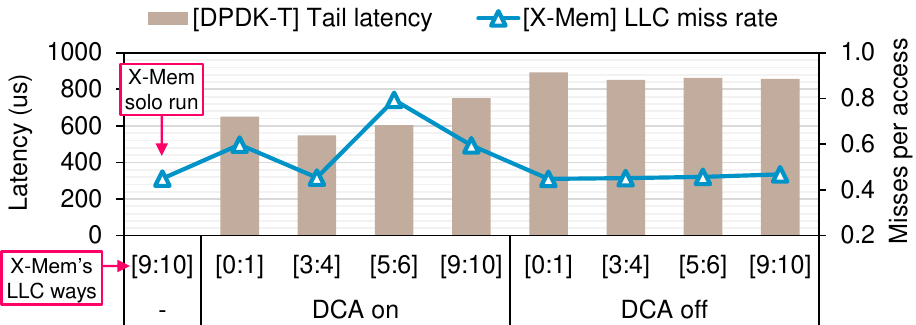}
     \vspace{-12pt}
     \caption{Validating the directory contention with DCA on.}
     \label{fig:dir_lat_cont}
     \Description{Bar/line chart comparing X‑Mem miss rate and DPDK‑T tail latency for four way placements with DCA on vs.\ off. Disabling DCA removes inclusive‑way contention but raises tail latency.}
     \vspace{-7pt}
\end{figure}

To validate this, we first disable DCA, forcing the CPU cores running DPDK-T to get all I/O cache lines through the device-memory-MLC path (\S\ref{subsec:ddio}).
Second, we allocate X-Mem to way[0:1] (DCA ways), way[3:4] (standard ways), way[5:6] (ways allocated to DPDK-T), and way[9:10] (inclusive ways), respectively.
\Fig{fig:dir_lat_cont} shows that disabling DCA avoids the contention between X-Mem and DPDK-T within inclusive ways because it does not bring I/O cache lines into MLCs from DCA ways. 
Note that disabling DCA also eschews the latent contention at DCA ways, but it considerably increases the p99 latency of DPDK-T (\S\ref{subsec:ddio_unfriendly_io}).
Based on this analysis, we make the following \textbf{O}bservation:

\vspace{0.4pt}
\aptLtoX{\begin{framed}
\noindent    \textbf{(O1)} When read by CPU cores, DMA-written I/O cache lines in DCA ways are migrated to inclusive ways, where I/O and non-I/O workloads contend.
\end{framed}}{
\noindent\fcolorbox{black}{white}{%
\minipage[t]{\dimexpr1\linewidth-2\fboxsep-2\fboxrule\relax}
    \textbf{(O1)} When read by CPU cores, DMA-written I/O cache lines in DCA ways are migrated to inclusive ways, where I/O and non-I/O workloads contend.
\endminipage}}

\subsection{Storage-IO-Driven DCA Contention}
\label{subsec:ddio_unfriendly_io}

\noindent\textbf{Setup.}
We use Flexible I/O Tester (FIO)~\cite{GitHubax76:online} as a storage-I/O workload, with four libaio (asynchronous I/O) threads.
Each libaio thread performs random read accesses to local storage-I/O devices, with the \texttt{O\_DIRECT} flag on (Direct I/O) and an I/O depth of 32.
They run on four CPU cores while allocated to way[2:3].
We modify FIO such that each thread also performs regular expression matching~\cite{regex:online} on storage blocks to make sure these blocks are brought into MLCs of the four CPU cores.  
Note that many datacenter servers deploy network-connected disaggregated storage servers but they still rely on local storage-I/O devices for fast access to performance-sensitive metadata or intermediate values~\cite{duggal2019data,klimovic2018understanding}.

\begin{figure}[!t]
    \vspace{-10pt}
    \captionsetup[subfigure]{justification=centering}
     \centering
 \subfloat[I/O throughout and memory bandwidth  
    \label{subfig:s3_leaky_dma_fio}]{%
       \includegraphics[page=1, width=0.73\columnwidth]{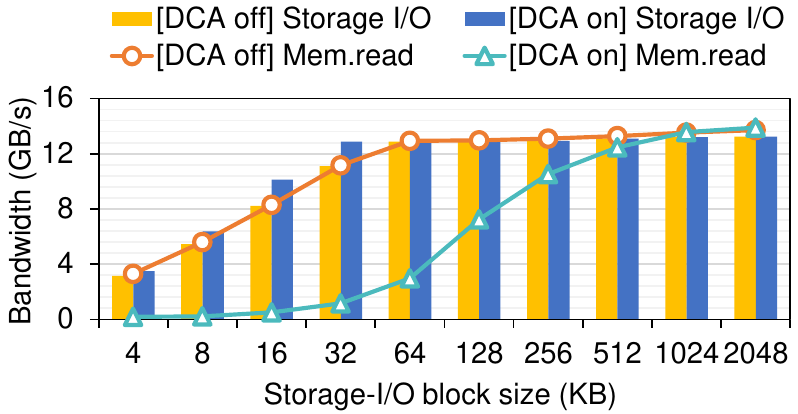}
     }
     \hfill
	\subfloat[DMA leak \label{subfig:s3_storage_dma_leak}]{%
       \includegraphics[page=1, width=0.23\columnwidth]{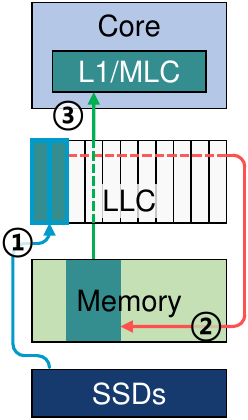}
     }
     \Description{Curves of storage‑I/O throughput and memory bandwidth as block size grows from 4KB to 2MB. Throughput plateaus beyond 128KB while bandwidth climbs.}
     \vspace{-6pt}
     \caption{Impact of storage block size and DCA on storage-I/O throughput, memory bandwidth, and DMA leak.}
     \label{fig:leaky_dma_fio}
     \Description{Step diagram illustrating DMA leak. SSD DMA writes into DCA ways, eviction of prior lines, CPU reads from memory, and write‑back cycle that causes leak.}
     \vspace{-6pt}
\end{figure}

\noindent\textbf{Storage-I/O characteristics.}
\label{subsec:ddio_unfriendly_io:stg_io}
\Fig{fig:leaky_dma_fio} shows the storage-I/O throughput and corresponding memory bandwidth consumption when DCA is enabled and disabled.
This demonstrates two important characteristics of storage I/O when the size of storage blocks is larger than 32KB. 
First, the storage-I/O throughput is little affected by whether DCA is enabled or disabled.
Second, although DCA is enabled, storage I/O consumes a considerable amount of memory bandwidth due to the DMA leak described below. 
As storage-I/O devices DMA-write large blocks to DCA ways at high rates (\raisebox{.4pt}{\textcircled{\raisebox{-.7pt} {1}}} in \Fig{subfig:s3_storage_dma_leak}), I/O cache lines, which stores previously DMA-written blocks, are evicted from these DCA ways before they are consumed by CPU cores (\raisebox{.4pt}{\textcircled{\raisebox{-.7pt} {2}}}).
Later, these I/O cache lines must be brought back from memory into MLCs (\raisebox{.4pt}{\textcircled{\raisebox{-.7pt} {3}}}).
We also observe that storage-I/O characteristics with employing deep I/O queues are similar to those with using large storage blocks.
Note that while we may mitigate the DMA leak caused by storage I/O by choosing smaller blocks and/or shallower queues, such an approach is not desirable for high-throughput storage-I/O applications~\cite{lee2021asynchronous,qian2024combining, bhattacharya2003asynchronous,tseng2021demystifying}.

Lastly, the effectiveness of DCA heavily depends on the I/O workload's temporal locality (\ie, the delay between I/O devices DMA-writing I/O data and CPU cores accessing/processing them).
Meanwhile, our experiment uncovers that storage-I/O-intensive workloads exhibit poor temporal locality for the following reasons.
Storage-I/O workloads often transfer one to three orders of magnitude more data per I/O transaction than network-I/O workloads.
This results in a longer transfer time per I/O transaction while CPU cores can start processing the data only after the entire I/O block has transferred to DCA ways.
Additionally, with more complex processing for larger amounts of data, it takes longer time for CPU cores to process the data, and subsequent DMA writes likely evict cache lines storing the data before all of it is accessed and processed (\ie, DMA leak).
Since our experiment with FIO assumes an optimistic scenario with minimal processing per block, other workloads may not benefit from DCA even at smaller block sizes.

\begin{figure}[!t]
    \captionsetup[subfigure]{justification=centering}
     \centering
 \subfloat[DPDK-T + FIO\label{subfig:dpdk_fio_interference:a}]{\includegraphics[page=1, width=0.73\columnwidth]{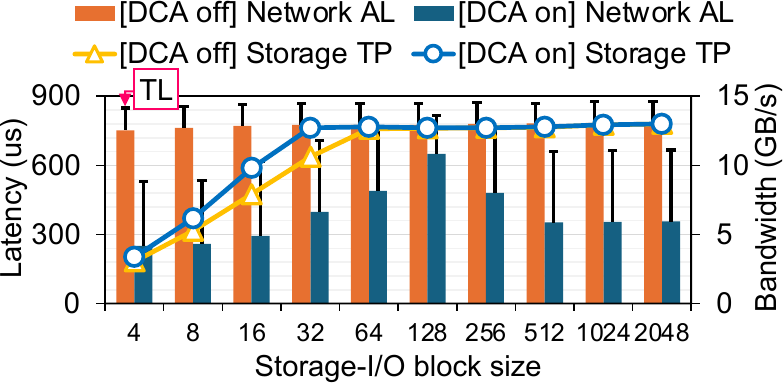}
     }
     \hfill
\subfloat[DPDK-T \label{subfig:dpdk_fio_interference:b}]{\includegraphics[page=2, width=0.23\columnwidth]{fig/motivation_graphs.pdf}
     }
     \vspace{-2.5pt}
     \caption{Impact of FIO on DPDK-T latency. `AL', `TL', and `TP' denote average latency, tail (p99) latency, and throughput.}
     \label{fig:dpdk_fio_interference}
     \Description{Paired bars of network latency of DPDK-T solo-run with and without DCA.}
    \vspace{-10pt}
\end{figure}

{\noindent \textbf{Contention between I/O workloads.}}
Noting that high-throughput storage I/O alone can incur a substantial amount of DMA leak, we hypothesize significant contention between storage- and network-I/O cache lines within DCA ways, as both can simultaneously DMA-write a large number of storage blocks and network packets to these DCA ways.
To demonstrate such contention, we co-run FIO with various storage block sizes and DPDK-T with a 2K-entry ring buffer per CPU core, and allocate them to way[2:3] and way[4:5], respectively. 
\Fig{fig:dpdk_fio_interference} shows that co-running DPDK-T with FIO increases the average latency of DPDK-T by 5--175\%, compared to running DPDK-T alone, when DCA is enabled. 
As the storage block size increases, the amount of DMA leak---and consequently, the latency of DPDK-T---also increases, peaking at a storage block size of 128KB, where storage-I/O throughput saturates. 
However, as the storage block size increases over 128KB, the latency of DPDK-T begins to decrease for the following reason.
In fact, we observe that the DMA leak occurs not only from DCA ways but also from inclusive ways, where migrated I/O cache lines are write-updated by subsequent DMA write (\Sec{subsec:directory_contention}) but evicted before consumed.
As the storage block size increases over 128KB, DMA leak begins to decrease, because 
most storage-I/O cache lines are evicted from DCA ways far before they are consumed and migrated to inclusive ways.
This makes most storage-I/O data brought into MLCs directly from the memory (\ie, no migration of storage-I/O cache lines to inclusive ways), which allows network-I/O data to use most of the capacity of inclusive ways.
To avoid the contention between FIO and DPDK-T within DCA and inclusive ways, we may choose to disable DCA, which does not affect the throughput of FIO. 
However, doing so unacceptably increases the latency of DPDK-T (\Fig{fig:dpdk_fio_interference}).

\vspace{0pt}
\aptLtoX{\begin{framed}
\noindent     \textbf{(O2)} Storage I/O can cause intense contention between storage-I/O and network-I/O cache lines within both DCA and inclusive ways, unacceptably increasing the latency of network I/O.

\end{framed}}{
\noindent\fcolorbox{black}{white}{%
\minipage[t]{\dimexpr1\linewidth-2\fboxsep-2\fboxrule\relax}
    \textbf{(O2)} Storage I/O can cause intense contention between storage-I/O and network-I/O cache lines within both DCA and inclusive ways, unacceptably increasing the latency of network I/O.
\endminipage}
}
\vspace{0.2pt}

\section{Strategies to Mitigate LLC Contentions}
\label{sec:mitigation}

In this section, we present strategies to mitigate the newly discovered LLC contentions, using both well-known and hidden knobs provided by modern server CPUs.

\subsection{Directory Contention-aware LLC Allocation}
\label{sec:mitigation:allocation}
We have shown that I/O workloads consuming I/O cache lines cause these cache lines to migrate to inclusive ways, evicting cache lines of workloads explicitly allocated to those ways, \ie, directory contention (\Sec{subsec:directory_contention}).  
We can avoid this contention by not allocating any workloads to inclusive ways using CAT, but at the cost of reducing the number of LLC ways available for workloads by almost 20\% (2 out of 11 ways). 
To address the inefficiency of such a na\"ive LLC allocation strategy, we propose explicitly allocating I/O workloads to LLC ways that encompass (or overlap) inclusive ways.

\begin{figure}[!t]
     \captionsetup[subfigure]{justification=centering}
     \centering
	\subfloat[Two LLC allocation strategies\label{subfig:alloc_strategy}]{\includegraphics[width=0.95\columnwidth]{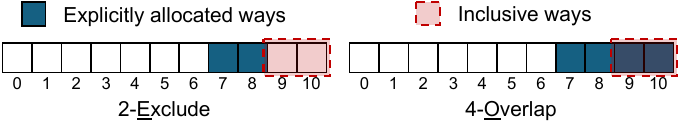}%
    }
    \hfill
     \vspace{4pt}
     \subfloat[Overlapping/Excluding\\inclusive ways\label{subfig:dpdk_diffway}]{\includegraphics[page=3, width=0.571\columnwidth]{fig/motivation_graphs.pdf}%
     }
     \Description{Diagram comparing n‑Exclude and n‑Overlap LLC allocations. Colored rectangles mark ways given to DPDK‑T, and dashed rectangles mark inclusive ways.}
     \hfill
	\subfloat[Life cycle of network Rx ring buffer\label{subfig:rx_life_cycle}]{\includegraphics[page=4,width=0.423\columnwidth]{fig/motivation_graphs.pdf}%
     }
     \Description{Bar graph of DPDK‑T latency and line graph of memory bandwidth for multiple Overlap/Exclude configurations. Overlap variants consistently show lower latency and bandwidth.}
     \hfill
     \vspace{-9pt}
     \caption{Impact of LLC allocation strategy on DPDK-T latency. `AL', and `TL' denote average and tail (p99) latency.}
     \label{fig:overlapping_ways}
     \Description{Lifecycle diagram of a network RX ring buffer depicting DMA write, CPU read, consumption, MLC eviction, and write‑update into inclusive ways.}
     \vspace{-12pt}
\end{figure}

Consider two LLC allocation strategies, \highlight{\textit{n}-Exclude} and \highlight{\textit{n}-Overlap}, where DPDK-T is explicitly allocated to \highlight{\textit{n}} LLC ways \highlight{E}xcluding and \highlight{O}verlapping inclusive ways, respectively.
\highlight{\textit{n}-Exclude} intends to completely avoid the directory contention with the na\"ive strategy, while \highlight{\textit{n}-Overlap} aims to maximize the LLC-caching efficiency.
For instance, \Fig{subfig:alloc_strategy} illustrates allocated ways of \highlight{2-Exclude} and \highlight{4-Overlap}.
Both \highlight{2-Exclude} and \highlight{4-Overlap} effectively use the same number of LLC ways, since DPDK-T with \highlight{2-Exclude} caches migrated I/O cache lines in inclusive ways.

\Fig{subfig:dpdk_diffway} shows that \highlight{(\textit{n+2})-Overlap} consumes less memory bandwidth (\ie, fewer LLC misses) and presents lower DPDK-T latency than \highlight{\textit{n}-Exclude} although both implicitly use the same number of LLC ways (green boxes) for two reasons.
First, for example, as \highlight{4-Overlap} and \highlight{2-Exclude} DMA-bloat 50\% and 100\% of consumed I/O cache lines back to way[7:8], respectively, \highlight{4-Overlap} incurs fewer conflict misses within way[7:8], thereby consuming less memory bandwidth.
Second, another 50\% of these I/O cache lines within the inclusive ways are soon reused/write-updated within those ways, which can provide these I/O cache lines for CPU cores more efficiently than I/O cache lines from other LLC ways for the following reason. 
Generally, \highlight{\textit{n}-Overlap} presents three types of I/O cache lines in inclusive ways (\Fig{subfig:rx_life_cycle}): {\raisebox{-0.3ex}{\scalebox{1.2}{\ding{108}}}\hspace{-1.8ex}\raisebox{0.1ex}{\textcolor{white}{1}}\hspace{0.6ex}} DMA-bloated (subset of dark blue), {\raisebox{-0.3ex}{\scalebox{1.2}{\ding{108}}}\hspace{-1.8ex}\raisebox{0.15ex}{\textcolor{white}{2}}\hspace{0.6ex}} write-updated (subset of yellow), and {\raisebox{-0.3ex}{\scalebox{1.2}{\ding{108}}}\hspace{-1.75ex}\raisebox{0.1ex}{\textcolor{white}{3}}\hspace{0.6ex}} migrated I/O cache lines (consumed ones and freed-after-consumed ones in subset of red and light blue).
Considering the life cycle of the network Rx ring buffer experiencing DMA bloat (\Fig{subfig:rx_life_cycle}): {\raisebox{.4pt}{\textcircled{\raisebox{-.7pt} {1}}}} DMA write, \raisebox{.4pt}{\textcircled{\raisebox{-.7pt} {2}}} CPU read, \raisebox{.4pt}{\textcircled{\raisebox{-.7pt} {3}}} CPU consumption, and \raisebox{.4pt}{\textcircled{\raisebox{-.7pt} {3}}} MLC eviction, we note that \raisebox{-0.3ex}{\scalebox{1.2}{\ding{108}}}\hspace{-1.8ex}\raisebox{0.15ex}{\textcolor{white}{1}}\hspace{0.6ex} is likely write-updated by DMA write and then becomes \raisebox{-0.3ex}{\scalebox{1.2}{\ding{108}}}\hspace{-1.8ex}\raisebox{0.1ex}{\textcolor{white}{2}}\hspace{0.6ex}. 
In contrast, \raisebox{-0.3ex}{\scalebox{1.2}{\ding{108}}}\hspace{-1.75ex}\raisebox{0.1ex}{\textcolor{white}{3}}\hspace{0.6ex} is migrated to inclusive ways after fresh I/O cache lines are write-allocated to DCA ways or DMA-bloated I/O caches lines write-updated within standard ways, both of which go through longer paths than {\raisebox{-0.3ex}{\scalebox{1.2}{\ding{108}}}\hspace{-1.8ex}\raisebox{0.15ex}{\textcolor{white}{1}}\hspace{0.6ex}} before CPU read.
Therefore, as the percentage of \raisebox{-0.3ex}{\scalebox{1.2}{\ding{108}}}\hspace{-1.8ex}\raisebox{0.15ex}{\textcolor{white}{1}}\hspace{0.6ex} in inclusive ways is higher with \highlight{\textit{n}-Overlap}, CPU cores can get I/O cache lines faster with fewer conflict misses within DCA ways.
We also observe analogous trends in LLC allocation strategy encompassing DCA ways.
%

\vspace{1pt}
\aptLtoX{\begin{framed}
    \textbf{(O3)} 
    The allocation of inclusive ways (and DCA ways) only to I/O workloads eases directory contention while maximizing LLC-caching efficiency.
\end{framed}}{
\noindent\fcolorbox{black}{white}{%
\minipage[t]{\dimexpr1\linewidth-2\fboxsep-2\fboxrule\relax}
    \textbf{(O3)} 
    The allocation of inclusive ways (and DCA ways) only to I/O workloads eases directory contention while maximizing LLC-caching efficiency.
\endminipage}}
\vspace{1pt}

\subsection{I/O Device-aware DCA and LLC Allocation}
\label{sec:mitigation:dcaoff}
To address storage-I/O-driven contention within DCA ways, we may disable DCA since DCA does not affect storage-I/O throughput  (\Fig{subfig:s3_leaky_dma_fio}) while only causing contention between storage- and network-I/O cache lines within DCA ways (\Fig{subfig:dpdk_fio_interference:a}). 
A well-known knob to disable DCA is through a BIOS option~\cite{disablingddio}. 
However, this option disables DCA for all I/O devices including network-I/O devices, which significantly increases the p99 of network-I/O workloads compared to enabling DCA (\Fig{subfig:dpdk_fio_interference:b}).

\begin{figure}[!t]
     \captionsetup[subfigure]{justification=centering}
     \centering
     \subfloat[DPDK-T + FIO\label{subfig:selective_dca}]{\includegraphics[page=5, width=0.641\columnwidth]{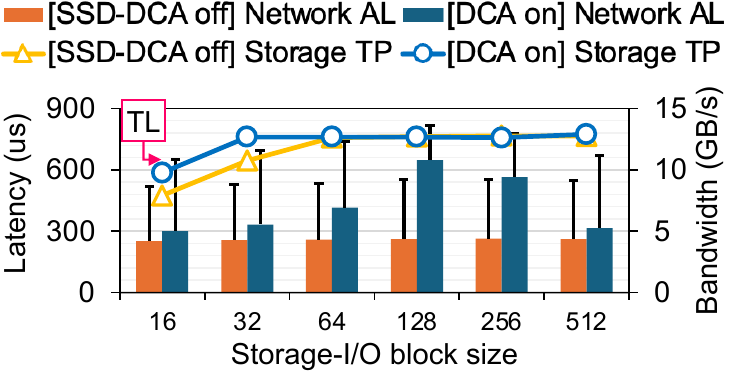}}
     \hfill
	\subfloat[FIO + X-Mem\label{subfig:storage_bloating}]{\includegraphics[page=6, width=0.351\columnwidth]{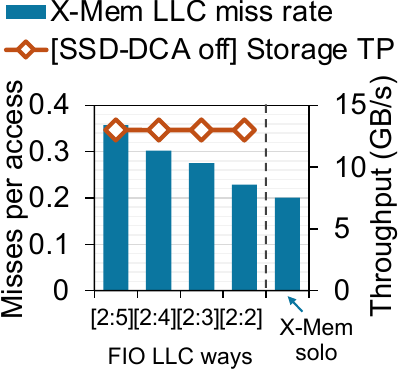}}
     \vspace{-9pt}
     \caption{Impact of I/O device-aware DCA disabling and LLC allocation on DPDK-T latency, FIO throughput, X-Mem LLC miss rate, where X-Mem is allocated to way[2:5].}
     \label{fig:device_aware_stategies}
     \Description{Bars of X‑Mem LLC miss rate and line of FIO throughput as FIO’s allocated standard ways shrink from four to one. Miss rate drops while throughput remains constant.}
     \vspace{-12pt}
\end{figure}

Tackling this challenge, we exploit a hidden feature that allows us to selectively disable DCA for a specific I/O device at runtime.
Specifically, we can disable DCA, setting \texttt{NoSnoopOpWrEn} and unsetting \texttt{Use\_Allocating\_Flow\_Wr} in register \texttt{perfctrlsts\_0} for each PCIe port~\cite{datasheetskx}.
Using this feature, we propose to disable DCA only for storage-I/O devices.
\Fig{subfig:selective_dca} plots the DPDK-T latency and FIO throughput when DCA for network-I/O devices is enabled but DCA for storage-I/O devices is disabled (\highlight{[SSD-DCA}~\highlight{off]}), demonstrating the effectiveness of \highlight{[SSD-DCA}~\highlight{off]} compared to when DCA is enabled for both network-I/O and storage-I/O devices (\highlight{[DCA}~\highlight{on]}). 
For storage blocks larger than 32KB, \highlight{[SSD-DCA}~\highlight{off]} offers 17--60\% and 18--29\% lower average and p99 latency for DPDK-T than \highlight{[DCA}~\highlight{on]}, respectively, while providing uncompromised throughput for FIO.   
That is, \highlight{[SSD-DCA~off]} practically eliminates the latency increase of DPDK-T caused by co-running FIO, compared to running DPDK-T alone, providing only 3\% and 9\% higher average and p99 latency, respectively.

However, with \highlight{[SSD-DCA}~\highlight{off]}, FIO DMA-bloats consumed storage-I/O cache lines back to other LLC ways, incurring contention within those ways.
To address such contention, we propose allocating as few standard ways as possible to storage-I/O workloads for the following reasons. 
The size of storage-I/O buffers often exceeds 10MB, which is far larger than that of network-I/O buffers and comparable to LLC. 
Therefore, consumed storage-I/O cache lines DMA-bloated back to standard ways cannot benefit from LLC caching as they will be evicted to the memory before they are reused/write-updated, unlike network-I/O cache lines. 
\Fig{subfig:storage_bloating} plots the LLC miss rate of X-Mem co-running with FIO, where X-Mem is allocated to way[2:5] and FIO is allocated to way[2:$n$], with $n$ gradually reduced from 5 to 2.
This shows that as $n$ decreases from 5 to 2, the LLC miss rate of X-Mem decreases from 36\% to 23\% due to fewer standard ways overlapped between FIO and X-Mem, while the throughput of FIO remains almost constant. 
So far, we have shown that storage-I/O workloads heavily pollute all LLC way by DMA-write and DMA bloat.
Furthermore, disabling DCA and restricting the LLC ways of such workload effectively addresses the pollution.
The effect on the storage-I/O workload is negligible as large block data have poor locality.
%

\vspace{0.7pt}
\aptLtoX{\begin{framed}
\noindent    \textbf{(O4)} Disabling DCA for storage-I/O devices eliminates DCA contention.
    \textbf{(O5)} Allocating only one standard way to storage-I/O workloads eases DMA bloat, without compromising storage-I/O throughput.
\end{framed}}{
\noindent\fcolorbox{black}{white}{%
\minipage[t]{\dimexpr1\linewidth-2\fboxsep-2\fboxrule\relax}
    \textbf{(O4)} Disabling DCA for storage-I/O devices eliminates DCA contention.
    \textbf{(O5)} Allocating only one standard way to storage-I/O workloads eases DMA bloat, without compromising storage-I/O throughput.
\endminipage}}
\section{A4: Share More, Interfere Less}
\label{sec:smartllc_overview}

\subsection{Overview}
\label{subsec:overview}

Leveraging the observations made thus far, we introduce \design, a holistic LLC management framework that orchestrates LLC resources among diverse co-running workloads.
Determining the optimal LLC allocation for individual workloads is not trivial, as their data access patterns and working set sizes heavily depend on the dynamic behavior of workloads and systems, such as I/O stack characteristics and OS paging. 
Furthermore, datacenter servers, typically equipped with more CPU cores than available LLC ways, enforce the sharing of LLC ways among multiple workloads, further complicating this allocation process.
To address these challenges, rather than statically allocating LLC ways to individual workloads, \design empirically and iteratively finds optimal LLC allocation by monitoring their relevant hardware performance counters.

\begin{figure}[!t]
\centerline{\includegraphics[width=0.95\columnwidth]{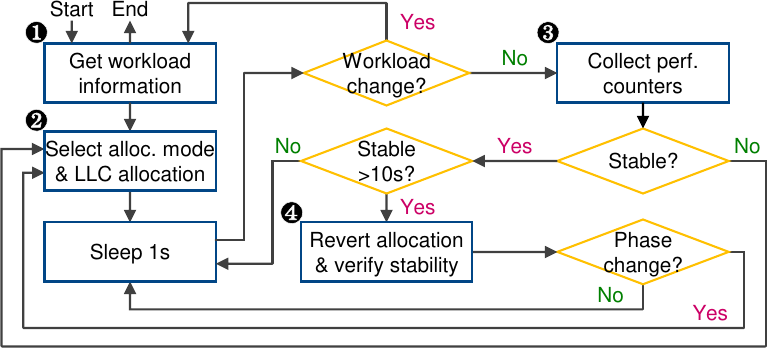}}
\vspace{-4pt}
\caption{\design execution flow.}
\label{fig:s6_smartllc_execution_flow}
\Description{A4 control‑loop flowchart. Workload info intake, performance‑counter sampling, mode selection, stability check with optional revert, then sleep cycle.}
\vspace{-12pt}
\end{figure}

\Fig{fig:s6_smartllc_execution_flow} shows the execution flow of \design.
When a workload is launched or terminated, \design gathers workload information, like QoS requirements (\ie, priorities) and associated I/O devices (\raisebox{-0.4ex}{\scalebox{1.2}{\ding{108}}}\hspace{-1.8ex}\raisebox{0ex}{\textcolor{white}{1}}\hspace{0.6ex}). 
This information can be provided by users or cluster management software~\cite{tirmazi2020borg,yuan2021don}.
Given the common practice in modern datacenters to co-locate workloads with varying priorities on the same server for better resource utilization~\cite{han2016interference,mars2011bubble}, \design finds optimal LLC allocations for workloads (\raisebox{-0.4ex}{\scalebox{1.2}{\ding{108}}}\hspace{-1.8ex}\raisebox{0ex}{\textcolor{white}{2}}\hspace{0.6ex}), considering (1) priority levels (\Sec{subsec:priority_llc_partition}), (2) contention between I/O and non-I/O workloads (\Sec{subsec:ddio_isolation}), (3) contention between network-I/O and storage-I/O workloads (\Sec{subsec:ddio_control}), and (4) other forms of contentions (\Sec{subsec:pseudo_llc_bypass}).
Subsequently, \design periodically monitors hardware performance counters to identify any antagonistic behavior or changes in workload execution phases (\raisebox{-0.4ex}{\scalebox{1.2}{\ding{108}}}\hspace{-1.75ex}\raisebox{0ex}{\textcolor{white}{3}}\hspace{0.6ex}, \Sec{subsec:cope_with_phase_change}).
For other phase changes that are harder to detect, if \design remains stable for 10 seconds, it reverts temporarily to the initial allocation to check if workloads are still in the same phase (\raisebox{-0.4ex}{\scalebox{1.2}{\ding{108}}}\hspace{-1.8ex}\raisebox{0ex}{\textcolor{white}{4}}\hspace{0.6ex}, \Sec{subsec:cope_with_phase_change}).
After \raisebox{-0.4ex}{\scalebox{1.2}{\ding{108}}}\hspace{-1.75ex}\raisebox{0ex}{\textcolor{white}{3}}\hspace{0.6ex} and \raisebox{-0.45ex}{\scalebox{1.2}{\ding{108}}}\hspace{-1.85ex}\raisebox{0ex}{\textcolor{white}{4}}\hspace{0.6ex}, \design initiates LLC reallocation if necessary.
Guidelines for tuning \design parameters are provided in \Sec{subsec:guidelines}.

\begin{figure*}[!t]
    \centering
    \includegraphics[width=\textwidth]{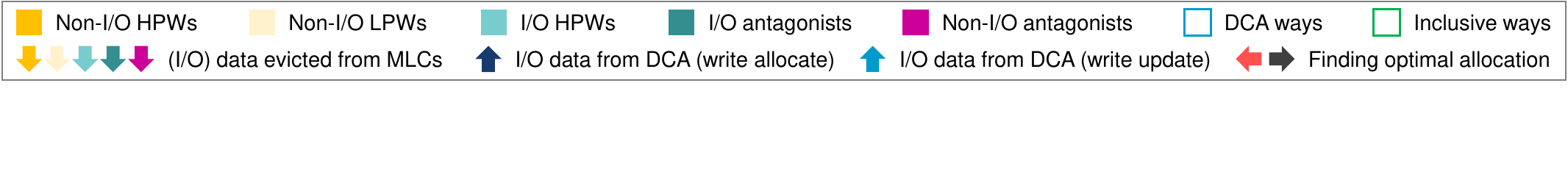}
    \vspace{-34pt}

     \captionsetup[subfigure]{justification=centering}
     \centering
\subfloat[Priority-based LLC allocation:\\non-I/O HPWs + non-I/O LPWs\label{subfig:s4_smartllc_example_1}]{\includegraphics[width=0.24\textwidth]{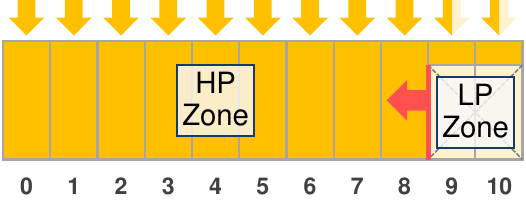}%
     }
     \Description{Diagram of priority‑based LLC partitioning: HP Zone initially owns all 11 ways, LP Zone starts with two right‑most ways.}
     \hfill
 \subfloat[Safeguarding I/O buffers:\\+ I/O-HPWs\label{subfig:s4_smartllc_example_2}]{\includegraphics[width=0.24\textwidth]{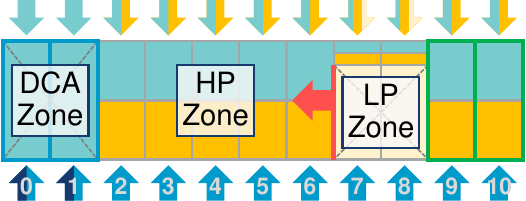}%
     }
     \Description{Diagram showing creation of DCA Zone (way[0–1]) and shift of LP Zone to avoid inclusive ways when I/O HPWs launch.}
     \hfill
 \subfloat[Disabling DCA:\\+ I/O antagonists\label{subfig:s4_smartllc_example_3}]{\includegraphics[width=0.24\textwidth]{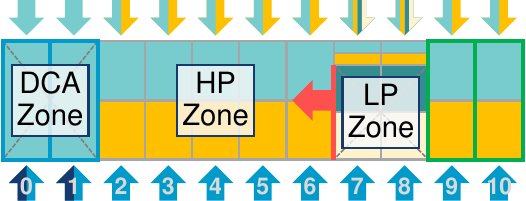}%
     }
     \Description{Diagram illustrating selective DCA disable for storage‑I/O workloads, relocating them into LP Zone while keeping DCA for network devices.}
     \hfill
 \subfloat[Pseudo LLC bypassing:\\+ non-I/O antagonists\label{subfig:s4_smartllc_example_4}]{\includegraphics[width=0.24\textwidth]{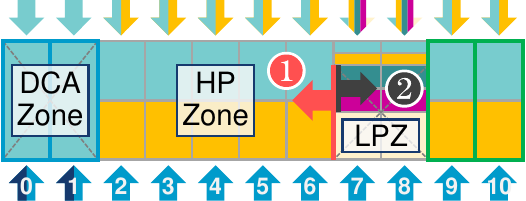}%
     }
     \Description{Diagram of pseudo‑LLC‑bypassing: DMA‑bloated and cache‑unfriendly workloads confined to minimal trash ways at right end of LLC.}
     \vspace{-4pt}
     \caption{\design operations with different workload combinations. We assume that, from left to right, different types of workloads are launched (+) one by one. Colored rectangles indicate LLC ways allocated to individual workload types (LPZ: LP Zone).}
     \label{fig:s4_smartllc_example}
     \vspace{-5pt}
\end{figure*}

\subsection{Priority-Based LLC Allocation}
\label{subsec:priority_llc_partition}

\design starts with allocating LLC ways based on the priorities of individual workloads (\Fig{subfig:s4_smartllc_example_1}): HP Zone shared for HPWs and LP Zone shared for LPWs. 
Without active I/O workloads, \design allocates all available LLC ways (way[0:10]) to HP Zone, while initially confining LP Zone to two LLC ways (way[9:10]), which we call \textit{initial partitions}. 
Then, \design expands LP Zone one LLC way at a time towards the left side (red arrow in \Fig{subfig:s4_smartllc_example_1}), aiming to prevent over-provisioning of the LLC resources to HPWs beyond their working set sizes while making the best effort for LPWs. 
To achieve this, \design monitors the LLC hit rates of individual HPWs every second and expands LP Zone every 2 seconds based on stable post-expansion values unless any HPWs exhibit a decrease in LLC hit rates surpassing the predefined threshold (\highlight{HPW\_LLC\_HIT\_THR}). Instead of LLC hit rates, alternative performance-related metrics, such as Instructions Per Cycle (IPC), can also be used based on specific Service-Level Objectives (SLOs).

\subsection{Safeguarding I/O Buffers}
\label{subsec:ddio_isolation} 
Once I/O HPWs are launched (\Fig{subfig:s4_smartllc_example_2}), \design intelligently rearranges LLC Zones to prevent contentions between I/O and non-I/O workloads.
Specifically, it exclusively reserves DCA ways (way[0:1]) as DCA Zone, allowing only I/O HPWs to allocate these LLC ways, to eliminate latent contention while maximizing LLC-caching efficiency for I/O buffers (O3 in \Sec{sec:mitigation:allocation}).
Meanwhile, it prevents LP Zone from allocating inclusive ways (way[9:10]) as well, since the cumulative eviction traffic generated by LPWs could incur significant directory contention (O1 in \Sec{subsec:directory_contention}).
That is, \design re-evaluates the optimal LLC allocation for LP Zone (as detailed in \Sec{subsec:priority_llc_partition}), starting with new initial partitions that designate way[7:8] for LP Zone and way[2:10] for HP Zone, while reserving way[0:1] for DCA Zone.

Note that \design still allocates inclusive ways as a part of HP Zone for three reasons.
First, HPWs utilize 9 LLC ways (way[2:10]), so their eviction traffic toward inclusive ways tends to be less intensive than that of LPWs which utilize fewer LLC ways and involve more antagonistic workloads (\Sec{subsec:pseudo_llc_bypass}).
Second, non-I/O HPWs can effectively utilize the remaining space in inclusive ways, preventing potential waste.
Inclusive ways are less crowded compared to DCA ways, as a subset of the I/O traffic is served directly by LLC (\ie, migrating to inclusive ways) while the rest is evicted to the memory and subsequently read through MLCs (\ie, DMA leak).
Third, even though some I/O data are evicted from inclusive ways by HPWs, most of them are already consumed and far from being reused, and thus, less critical than those in DCA ways.

\subsection{Disabling DCA to Prevent DMA Leak}
\label{subsec:ddio_control}

Intensive storage I/O can cause significant contention between storage-I/O and network-I/O cache lines within both DCA and inclusive ways (O2 in \Sec{subsec:ddio_unfriendly_io}).
Based on this observation, after launching storage-I/O workloads, \design monitors their I/O behavior and, upon detecting storage-I/O-driven contention, temporarily disables DCA for SSDs to mitigate DMA leak (O4 in \Sec{sec:mitigation:dcaoff}).
Furthermore, such a storage-I/O workload is treated as an LPW even if it was initially set to HPW, to mitigate pollution in HP Zone caused by heavy MLC-eviction traffic generated by storage-I/O data  (\ie, DMA bloat). 
At this point, LP Zone is reallocated (as detailed in \Sec{subsec:priority_llc_partition}), now including DCA-disabled storage-I/O workloads.

\design dynamically detects storage-I/O-driven contention by monitoring: 
(1) frequent eviction of I/O cache lines from the LLC, \ie, DCA miss rate surpasses \highlight{DMALK\_DCA\_MS\_THR}; 
(2) significant DMA leak, \ie, average LLC miss rate of storage-I/O workloads surpasses \highlight{DMALK\_LLC\_MS\_THR}; and
(3) substantial contribution of storage I/O to the DMA leak, \ie, storage-I/O portion in overall system I/O's read throughput (\aka PCIe write throughput) exceeds \highlight{DMALK\_IO\_TP\_THR}.
These collectively suggest storage I/O is causing considerable DMA leak without deriving any benefits from DCA. Therefore, we can safely decide to disable DCA.

\subsection{Pseudo LLC Bypassing}
\label{subsec:pseudo_llc_bypass}

DCA-disabled storage-I/O workloads extensively read/write low-locality (I/O) data from/to the memory and cause contention within LP Zone.
Similar behavior is observed in memory-intensive, non-I/O workloads accessing data with poor temporal locality, such as KSM~\cite{stevenson2013fine} and zswap~\cite{weiner2022tmo} that are widely used for system resource management.
To alleviate these contentions, we introduce pseudo LLC bypassing (\Fig{subfig:s4_smartllc_example_4}), which manages both DMA-bloated and cache-unfriendly cache lines together by allocating as few standard ways as possible to such workloads (O5 in \Sec{sec:mitigation:dcaoff}).
After LP Zone (LPZ) settles (\raisebox{-0.45ex}{\scalebox{1.2}{\ding{108}}}\hspace{-1.75ex}\raisebox{0ex}{\textcolor{white}{1}}\hspace{0.6ex}), \design monitors the MLC and LLC miss rates of each non-I/O workload.
If both metrics exceed \texttt{ANT\_CACHE\_MISS\_THR}, \design speculates that this workload is an antagonist that derives minimal benefit from LLC caching. 
Such a non-I/O antagonist is also treated as LPW.
For all identified I/O and non-I/O antagonists, \design progressively reduces the allocated LLC ways (\ie, trash ways) from the current LP Zone down to the rightmost standard way (way[8], \raisebox{-0.5ex}{\scalebox{1.2}{\ding{108}}}\hspace{-1.8ex}\raisebox{0ex}{\textcolor{white}{2}}\hspace{0.6ex}), directing (DMA-bloated) dead cache lines to a subset of the LLC ways allocated to LP Zone.
To prevent drastic performance drop, memory bandwidth abuse, or incorrect antagonist detection, this process ceases if any of the following exhibit instability (\eg, fluctuations greater than 10\%) after LLC way reductions: (1) LLC miss rates of non-I/O antagonists, (2) I/O throughput of storage-I/O antagonists, or (3) system-wide memory bandwidth.

Note that changing LLC way affinity using CAT only affects newly allocated LLC lines.
Therefore, we can alleviate the LLC contention by selectively applying pseudo LLC bypassing to a workload only when it exhibits antagonistic behavior, while still maintaining its performance-critical data in excluded LLC ways.

\subsection{Reacting to Execution Phase Changes}
\label{subsec:cope_with_phase_change}

Once the LLC allocation is complete (\Sec{subsec:priority_llc_partition}, \Sec{subsec:ddio_isolation}, \Sec{subsec:ddio_control}, \Sec{subsec:pseudo_llc_bypass}), \design performs continuous, per-second monitoring of each workload's key performance metrics to cope with any detected changes in workloads and execution phases, as follows.

\niparagraph{Reallocating LP Zone.}  
\design reallocates LP Zone under the following three conditions:
(1) new HPW combinations: occurs at the launch or termination of HPWs, or active workload's transition between HPW and LPW states (\Sec{subsec:ddio_control}, \Sec{subsec:pseudo_llc_bypass}); 
(2) execution phase changes: triggered if any HPW shows a fluctuation in LLC hit rate exceeding \highlight{HPW\_LLC\_HIT\_THR} compared to that recorded in the initial partitions; and
(3) uncapturable changes: activated if any HPW exhibits a fluctuation in LLC hit rate exceeding \highlight{HPW\_LLC\_HIT\_THR} compared to the highest attainable hit rate at the moment, estimated by temporarily reverting to the initial allocation at every 10 seconds in a stable state. 

\niparagraph{Re-assigning priorities.} 
\design detects antagonistic workloads either after their launch (for HPWs) or when they reach a stable state (for both HPWs and LPWs). 
However, their behavior may vary over time, such as when antagonistic data accesses complete.
To adapt to these changes, \design restores LLC allocation for non-I/O antagonists if fluctuations appear in their LLC miss rate compared to that observed during initial antagonistic detection. For those originally classified as LPWs, the LLC allocation is reverted to the current LP Zone. 
Otherwise, they return to the HPW pool, triggering priority-based LLC reallocation.
For storage-I/O antagonists, a significant fluctuation in storage-I/O throughput signals major phase changes, prompting a restoration of LLC allocations based on initial QoS requirements and reactivating DCA for the SSDs.

\subsection{Guidelines for Tuning A4 Parameters}
\label{subsec:guidelines}
 
\design employs five threshold values (T1---T5 in \Tab{tab:config-params}), and two timing parameters.
Fine-tuning these values is crucial for maximizing the efficacy of \design, aligning with the needs of users and service providers.
Below, we offer guidelines for setting these values.

\niparagraph{T1.} 
This threshold determines the extent to which LP Zone can expand, balancing it against the effective capacity available to HP Zone.
With a lower threshold, HPWs could benefit from lower LLC miss rates, whereas a higher threshold allows LPWs to utilize more LLC resources.
Note that \design can be readily extended in coordination with existing system monitoring tools~\cite{yuan2021don, yang2013bubble, nathuji2010q, qureshi2006utility} to dynamically adjust this threshold according to specific SLOs.
For instance, it can begin with the lowest value and then gradually increase the threshold as long as HPWs' specific SLOs remain satisfied.

\niparagraph{T2--T4.} 
T2 and T3 are used together to detect whether I/O workloads suffer from DMA leak,
while T4 specifically determines whether storage I/O is the source of the DMA leak.
These threshold values should be configured considering the I/O devices attached to the system, their relative bandwidth, and the extent to which the DMA leak impacts the workloads utilizing these devices.
With lower thresholds, storage-I/O workload may be mistakenly classified as an antagonist.
However, \design provides a correction mechanism for misclassifications, allowing these parameters to be set aggressively.

\niparagraph{T5.} 
This threshold is used to identify potential non-I/O antagonists.
The threshold should be set sufficiently high to ensure the workload gains no benefit from using the LLC.
A low threshold may force the workload to sacrifice its performance for others.

\noindent \textbf{Timing parameters.}
 \design reverts to the initial partition for 1 second (revert interval) at every 10 seconds (stable interval) of a stable state (\S\ref{subsec:cope_with_phase_change}).
 Tuning a stable interval has a trade-off between responsiveness to an uncapturable phase change and performance.
 A short stable interval will result in frequent revert to the initial partition which yields lower performance gain.
 \design measures hit rates for a sufficiently long revert interval.
 Averaging the counter values for a long time ensures that \design to obtain stable values from workloads that frequently, and repeatedly change their phases.
 This guarantees that \design avoids frequent searches for the optimal allocation, except when an actual phase change occurs.
\section{Experimental Methodology}
\label{sec:methodology}

\begin{table}[!t]
\caption{\design evaluation setup.}
\vspace{-10pt}
\label{tab:config-params}
\begin{center}
\resizebox{\columnwidth}{!}{%
\begin{tabular}{|c||c|}
\hline 
\multicolumn{2}{|c|}{\rule[-1.0ex]{0pt}{3.4ex}\textbf{Server machine}}\\
\hline
OS (kernel)                         & Ubuntu-20.04.6 LTS (5.15.0-91-generic)\rule[-1.0ex]{0pt}{3.4ex}\\
\hline
\multirow{2}{*}{CPU}                & Intel\textsuperscript{\textregistered} Xeon\textsuperscript{\textregistered} Gold 6140 CPU @2.30 GHz, 18 cores, 32KiB private L1 \rule{0pt}{2.4ex}\\
& I/D\$, 1MiB private L2\$, 25MiB shared LLC (11 ways, non-inclusive)\rule[-1.0ex]{0pt}{0pt}\\
\hline
Main memory                         & 16GB, 32GB DDR4 DRAM per channel (Total: 288GB, 6 channels)\rule[-1.0ex]{0pt}{3.4ex} \\
\hline
Network device                      & 100Gbps Nvidia ConnextX-6 NIC (part of the BlueField-2 SoC)\rule{0pt}{2.4ex}\\
Storage device                      & 4$\times$ Samsung 980 PRO 1TB M.2 SSDs (4TB, PCIe Gen~3 $\times$16)\rule[-1.0ex]{0pt}{0pt}\\
\hline 
\multicolumn{2}{|c|}{\textbf{Client machine (network packet generator)}}\rule[-1.0ex]{0pt}{3.4ex}\\
\hline
OS (kernel)                         & Ubuntu-20.04.5 LTS (5.15.0-78-generic)\rule[-1.0ex]{0pt}{3.4ex}\\
\hline
CPU                                 & Intel\textsuperscript{\textregistered} Xeon\textsuperscript{\textregistered} E5-2650 v4 @2.20GHz, 12 cores, 30MiB LLC (20 ways)\rule[-1.0ex]{0pt}{3.4ex}\\
\hline
Main memory                         & 16GB DDR4 DRAM (Total: 16GB, 1 channel)\rule[-1.0ex]{0pt}{3.4ex}\\
\hline
Network device                      & 100Gbps Nvidia ConnextX-6 NIC (part of the BlueField-2 SoC)\rule{0pt}{2.4ex}\\
\hline 
\hline 
\multicolumn{2}{|c|}{\textbf{\design threshold values}}\rule[-1.0ex]{0pt}{3.4ex}\\
\hline 
\multicolumn{2}{|c|}{1. \texttt{HWP\_LLC\_HIT\_THR}: 20\%, 2. \texttt{DMALK\_DCA\_MS\_THR}: 40\%, 3. \texttt{DMALK\_IO\_TP\_THR}: 35\%}\rule{0pt}{2.4ex}\\
\multicolumn{2}{|c|}{4. \texttt{DMALK\_LLC\_MS\_THR}: 40\%, 5. \texttt{ANT\_CACHE\_MISS\_THR}: 90\%}\rule[-1.0ex]{0pt}{0pt}\\
\hline
\end{tabular}
}
\end{center}
\end{table}

\begin{table}[!t]
\vspace{-5pt}
\caption{Real-world workloads for evaluation.}
\vspace{-10pt}
\label{tab:config-apps}
\begin{center}
\resizebox{\columnwidth}{!}{%
\begin{tabular}{|c||c|c|}
\hline 
\textbf{Name} & \textbf{Description} & \textbf{Parameters}\rule[-1.0ex]{0pt}{3.4ex}\\
\hline
\multirow{2}{*}{Fastclick~\cite{tbarbett20:online}} & Network I/O workload & 1024-B packets, 2048-entry \rule{0pt}{2.4ex}
\\ & Simple packet processing  & ring buffer per CPU core, 4 CPU cores \rule[-1.0ex]{0pt}{0pt}\\
\hline
FFSB-H (heavy) and  & Storage I/O workload & Heavy: 2MB I/O block, 3 CPU cores \rule{0pt}{2.4ex}\\
FFSB-L (light)~\cite{FFSBPrim6:online} & Regular expression matching & Light: 32KB I/O block, 1 CPU core \rule[-1.0ex]{0pt}{0pt}\\
\hline
Redis-S (server) and & In-memory database workload & YCSB workload A (Update heavy), \rule{0pt}{2.4ex}\\
Redis-C (client)~\cite{Redis81:online} & Persistent key-value store & 1 CPU core each~\cite{cooper2010benchmarking} \rule[-1.0ex]{0pt}{0pt}\\
\hline
\multirow{2}{*}{SPEC CPU2017~\cite{spec2017}} & \multirow{2}{*}{General-purpose workloads} & SPECrates with reference\rule{0pt}{2.4ex}\\
&  &  input sets, 1 CPU core each\rule[-1.0ex]{0pt}{0pt}\\
\hline

\end{tabular}
}
\end{center}
\end{table}

\begin{table}[t]
\centering
\vspace{-5pt}
\caption{X-Mem instances for microbenchmark evaluation.}
\vspace{-10pt}
\label{tab:xmem-inst}
\begin{center}
\resizebox{\columnwidth}{!}{%
\begin{tabular}{|c|c|c|c|c|}
\hline
\textbf{Instance} & \textbf{Working set size (MB)} & \textbf{Access pattern} & \textbf{Operation type} \\ \hline
X-Mem 1       & 4MB                            & Sequential              & Read                 \\ \hline
X-Mem 2       & 4MB                           & Sequential                  & Write                 \\ \hline
X-Mem 3       & 10MB                            & Random                   & Read               \\ \hline
\end{tabular}
}
\end{center}
\vspace{-10pt}
\end{table}

{\noindent \textbf{Testbeds.}}
\Tab{tab:config-params} outlines our testbeds. The server machine serves as our focal system for LLC orchestration, running all the (micro)benchmarks along with a \design daemon. The client machine generates and sends network packets to the server by running DPDK Pktgen~\cite{ThePktge25:online,GitHubpk95:online}.
Two machines are connected via a 100Gbps network interface, driven by Nvidia ConnextX-6 NICs. 
We verified that Pktgen generates sufficient network traffic to stress the server, reaching up to 100Gbps.
The server's local storage consists of a RAID-0 array of four 1TB SSDs plugged into a PCIe Gen3 $\times$16 slot.
Hyper-Threading and Turbo Boost are disabled on both machines.

\noindent \textbf{System monitor/control.}
 Hardware performance counters, such as LLC/DCA hits/misses and I/O throughputs, are monitored using Intel Performance Counter Monitor (PCM)~\cite{intelpcm30:online}.
We use Intel CAT~\cite{GitHubin14:online} to allocate LLC way(s) to individual CPU cores.
Note that CAT only allows contiguous LLC way allocation for each CPU core.
\design's fundamental operations (system monitor and cache control) closely resemble those in \cite{yuan2021don}, which was confirmed to add negligible overhead to the system even with context switches for accessing hardware performance counters. 
We also verify that the execution time of \design does not exceed 800$\mu$s, imposing minimal overhead on the system given the 1-second monitoring interval.

{\noindent \textbf{Benchmarks.}}
We evaluate \design using both microbenchmarks (\Sec{subsec:microbench}) and real-world workloads (\Sec{subsec:real_world}).
For microbenchmarks, we use the same workloads described in \Sec{sec:motivation} but with varying configurations to offer a comprehensive analysis of \design's efficacy.
The real-world workloads are described in \Tab{tab:config-apps}.
Workloads are pinned to specific CPU cores, while \design daemon occupies one CPU core, ensuring its isolation from other CPU cores while remaining within the same NUMA node.
Result values are averaged over five iterations, and each of them lasts 70s, including the initial 10s of warm-up time and the last 10s of result-collecting time.

{\noindent \textbf{Evaluated designs.}}
We compare \design with two baseline LLC management schemes:
(1) a Default model, where all workloads share the entire LLC without explicit CAT allocation; and (2) an Isolate model, which statically assigns distinct LLC ways to each workload in proportion to the number of pinning CPU cores, \ie, static workload-wise LLC isolation.
In both baseline models, DCA is enabled for every I/O device.

\section{Results and Analysis}
\label{sec:results_analysis}

\subsection{Microbenchmark Results}
\label{subsec:microbench}

{\noindent \textbf{Setup.}
Besides DPDK-T (HPW) and FIO (LPW), each running on four CPU cores (way[2:3] and way[4:6], respectively, in the Isolate model), we run three X-Mem variants as depicted in \Tab{tab:xmem-inst}.
We conduct experiments with (1) varying network packet sizes from 64B to 1,514B with 2MB storage I/O blocks, and (2) varying storage I/O block sizes from 4KB to 2MB with 1,514B packets.
}

\begin{figure}[!t]
    \includegraphics[page=1, width=\columnwidth]{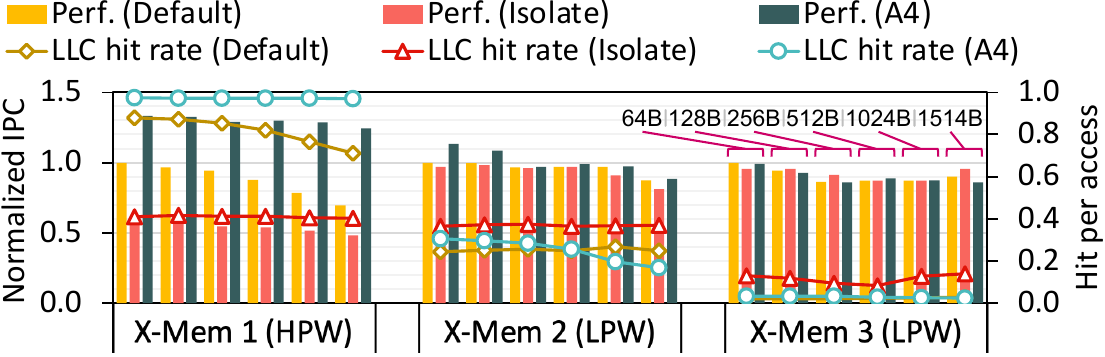}%
    \vspace{-6pt}
     \caption{IPC and LLC hit rates of three X-Mem variants with varying network packet sizes (storage I/O size: 2MB).}
     \label{fig:micro-xmem}
     \vspace{-10pt}
     \Description{Bars of normalized IPC and lines of LLC hit rate for three X‑Mem variants as network packet size scales. A4 line maintains high hit rate and IPC across sizes.}
\end{figure}

{\noindent \textbf{Non-I/O workload analysis.}}
\Fig{fig:micro-xmem} depicts the IPCs and LLC hit rates of three X-Mem variants with varying network packet sizes. 
IPC values are normalized to those of the Default model with the smallest packet size.  
In the Default model, three X-Mems show degraded performance with increasing packet sizes, because larger packets pose more pressure on the LLC as the I/O buffer size increases (\ie, DMA bloat).
On the other hand, in the Isolate model, their performance seems to be less affected by the packet sizes since LLC ways allocated to each workload are isolated from each other. 
Nevertheless, this performance isolation is imperfect because it can be easily broken by the directory and latent contentions. 
As a result, X-Mem 1 and X-Mem 2 show performance degradation with larger packets.
In addition, as it statically allocates LLC ways without considering cache sensitivity and working set sizes, a cache-sensitive X-Mem 1 occupies less effective LLC capacity than the Default model, resulting in lower performance.
In \design, X-Mem 1 (HPW) and X-Mem 2 (LPW) maintain their initial QoS requirements, while X-Mem 3 is detected as an antagonist.
As a result, X-Mem 1 achieves speedups of 1.3$\times$--1.78$\times$ over the Default model by maintaining consistent LLC hit rates of 97\% across varying packet sizes. 
Meanwhile, LPWs (X-Mem 2 and X-Mem 3) retain the performance within acceptable ranges.

\begin{figure}[!t]
    \includegraphics[page=2, width=\columnwidth]{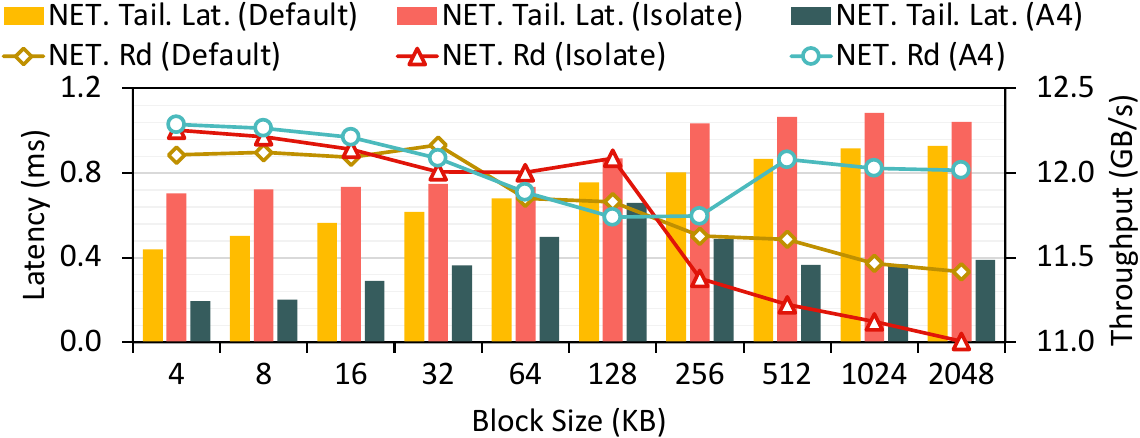}%
    \vspace{-6pt}
    \caption{Network performance metrics with varying storage I/O block sizes (network packet size: 1,514B).} 
    \label{fig:net-blk}
    \Description{Plots of network latency (left axis) and throughput (right axis) against storage block size. A4 flattens latency curve compared to Default and Isolate.}
    \vspace{-10pt}
\end{figure}

{\noindent \textbf{I/O workload analysis.}}
\Fig{fig:net-blk} depicts the average latency and throughput of DPDK-T with varying storage I/O block sizes. 
With increasing I/O block sizes, the Default and Isolate models exhibit increasing network latency and decreasing network throughput due to the exacerbated storage-I/O-driven contentions.
These trends are more pronounced in the Isolate model, as network I/O can benefit from reusing DMA-bloated I/O buffers only within smaller isolated LLC ways.
On the other hand, \design achieves a 58\% reduction in network latency and a 5\% increase in network throughput at the largest block size compared to the Default model.
Note that \design experiences gradual network performance degradation until the block size reaches 128KB, as FIO is not detected as an antagonistic workload within this range.

\begin{figure*}[!t]
    \captionsetup[subfigure]{justification=centering}
     \centering
     \vspace{-2pt}
     \subfloat[HPW-heavy scenario: 7 HPWs + 4 LPWs (3 antagonists detected). Speedups: 52\% (HPWs), 22\% (All) \label{subfig:hpw_heavy}]{\includegraphics[width=0.96\textwidth]{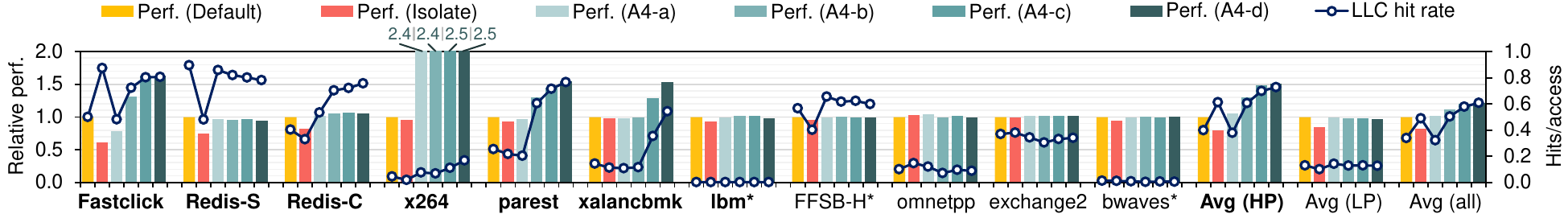}%
     }
     \Description{Bars of speedup and line of LLC hit rate for HPW‑heavy workload mix across orchestration schemes. A4‑d yields highest bars for HPWs without hurting LPWs.}
     \vspace{3pt}
     \hfill
     \subfloat[LPW-heavy scenario: 4 HPWs + 8 LPWs (4 antagonists detected). Speedups: 47\% (HPWs), 33\% (All) \label{subfig:lpw_heavy}]{\includegraphics[width=0.95\textwidth]{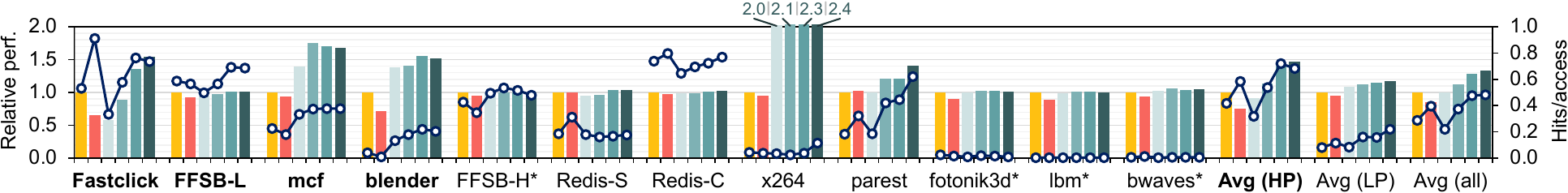}%
     }
     \Description{Similar graph for LPW‑heavy mix. A4 improves both HPWs and LPWs performance.}
     \vspace{-10pt}
     \caption{Performance and LLC hit rates of real-world workloads with various LLC orchestration schemes. Workloads with an original QoS requirement of \textbf{HPW} are in bold, and asterisks (*) indicate workloads experiencing pseudo LLC bypassing.}
     \label{fig:perf_llc_hit_rate}
\end{figure*}

\subsection{Real-World Workload Results}
\label{subsec:real_world}

Besides Default and Isolate models, we evaluate four variants of \design, applying the techniques proposed in \Sec{sec:smartllc_overview} (\Fig{subfig:s4_smartllc_example_1}--\ref{subfig:s4_smartllc_example_4}) one by one to Default model (denoted as \design-a--d, respectively).

\noindent \textbf{HPW-heavy scenario.} 
\Fig{subfig:hpw_heavy} shows the relative performance (normalized to Default model) and LLC hit rates of co-running workloads listed in \Tab{tab:config-apps}, with seven workloads assigned high priority and four assigned low priority.
We denote HPWs as boldface and LPWs as normal legends, and the workloads detected as antagonists are highlighted with asterisks.
We observe that IPCs and LLC hit rates do not always align with actual performance in multi-threaded I/O workloads.
This is because these metrics are often inflated by synchronization, I/O syscall, and idle loops~\cite{alameldeen2006ipc}.
For instance, LLC hit rates of Fastclick in the Isolated model tend to be inflated even though they exhibit lower throughput and higher latency compared to Default model.
Thus, we measure throughput (inverse of latency per request) for multi-threaded I/O workloads (Fastclick and FFSB-H/L) and IPC for single-threaded workloads (Redis-S/C and SPEC CPU).
These metrics are inversely proportional to the execution time, accurately representing their performance.

In general, Isolate model yields lower performance than Default model.
Isolate model confines each worklaod to specific LLC ways, irrespective of its QoS requirement and cache sensitivity.
Such rigid allocation impedes flexible LLC sharing among workloads, often resulting in mismatches where their working set sizes either exceed or fall short of the assigned LLC capacity.
Even with LLC sharing enabled among workloads of the same priority, \design-a (\Fig{subfig:s4_smartllc_example_1}) fails to demonstrate a significant performance improvement, except for x264.
This is primarily due to I/O-driven contentions, as both network-I/O and storage-I/O devices inject high-bandwidth I/O traffic into the LLC.
The I/O cache lines from these workloads, distributed across all LLC ways, significantly interfere with co-running non-I/O workloads, offsetting the benefit of priority-based partitioning.
Conversely, the accumulated MLC eviction traffic from LPWs to LP Zone (including inclusive ways) creates significant contention with Fastclick, leading to a noticeable performance degradation.
Nevertheless, FFSB-H is less affected by these contentions as its heavy storage I/O operations hardly benefit from DCA and LLC caching.
By safeguarding I/O buffers, \design-b (\Fig{subfig:s4_smartllc_example_2}) enhances the LLC hit rate and performance of Fastclick by 50\% and 67\%, respectively, compared to \design-a.
Non-I/O HPWs also gain from improved LLC-caching efficiency for I/O buffers, as a substantial portion of I/O cache lines now resides in either DCA or inclusive ways, alleviating DMA bloat to standard ways.
Consequently, \design-b demonstrates 32\% (29\%) performance improvements for I/O (non-I/O) HPWs and an overall 52\% increase in LLC hit rates across all HPWs compared to Default model.

Even with these optimizations, Fastclick still suffers from DMA leak and bloat caused by FFSB-H.
\design-c (\Fig{subfig:s4_smartllc_example_3}) intelligently disables DCA for FFSB-H by detecting its antagonistic behavior, resulting in a 11\% improvement in LLC hit rate and a 24\% performance increase for Fastclick compared to \design-b.
However, this increases the eviction traffic from its MLCs to LP Zone as FFSB-H can no longer utilize DCA and inclusive ways.
\design-d (\Fig{subfig:s4_smartllc_example_4}) relieves such DMA-bloated pressure on LP Zone, by directing the consumed I/O cache lines to limited trash ways.
As bwaves and lbm are also identified as an antagonist due to their high MLC/LLC miss rates, \design-d further improves the performance and LLC hit rates of HPWs by 2\% and 4\% compared to \design-c.
After mitigating latent, directory, and DCA contention, I/O HPW performance relies less on standard ways, yielding only marginal gains for \design-d over \design-c.

The sensitivity of non-I/O HPWs performance to LLC capacity is explained by the analysis of SPEC CPU2017 Suite~\cite{singh2019memory}.
It suggests that x264 reaches diminishing returns beyond a certain cache size, but parest and xalancbmk steadily benefit from increased cache size.
This aligns with our result that the LLC hit rates of parest and xalancbmk increase progressively, unlike x264.
To sum up, \design successfully eradicates various types of I/O-driven contentions.
%
\design improves LLC hit rate and performance of all (HPW) workloads by 79\% (83\%) and 22\% (51\%), respectively, compared to the Default model, without compromising the performance of LPWs.

\noindent \textbf{LPW-heavy scenario.} 
We verify the robustness of \design with a different combination of workloads, more focused on LPWs, described in ~\Fig{subfig:lpw_heavy}.
While the overall trend is similar to the HPW-heavy scenario, the LPW-heavy scenario demonstrates distinct trends emerging from a smaller HP Zone and a larger LP Zone.
First, non-I/O HPWs performance saturates right after applying \design-a.
Second, non-antagonistic LPWs, x264 and parest, benefit from extensive LP Zone capacity.
However, their performance gains are lower than in the HPW-heavy scenario where they are set as HPWs.
Note that for x264, while the LLC hit rate gradually increases, the L2 hit rate increases sharply.
In addition, \design distinguishes antagonistic workload among multiple storage I/O workloads, and selectively disables DCA for FFSB-H only.
Overall, \design improves LLC hit rate and performance of all (HPW) workloads by 68\% (63\%) and 33\% (47\%), respectively, compared to the Default model.

\begin{figure*}[!t]
     \vspace{-4pt}
    \captionsetup[subfigure]{justification=centering}
     \centering
     \subfloat[Breakdown of network I/O (Fastclick) average latency \label{subfig:net_brkdwn}]{\includegraphics[width=0.49\columnwidth]{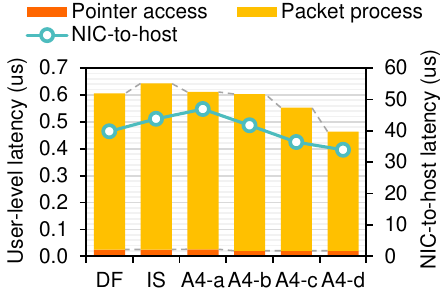}%
     }
     \Description{Stacked bars breaking Fastclick average latency into NIC‑host transfer, pointer access, packet processing. A4‑d shortens all segments versus baselines.}
     \hfill
     \subfloat[Breakdown of storage I/O (\textbf{F}F\textbf{S}B-\textbf{H}) average latency\label{subfig:stg_brkdwn}]{\includegraphics[width=0.49\columnwidth]{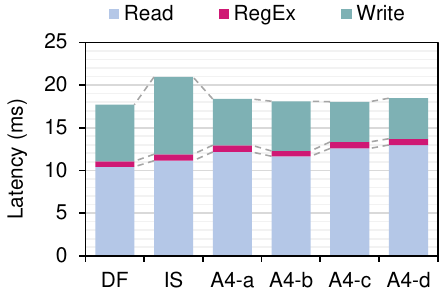}%
     }
     \Description{Stacked bars splitting FFSB‑H latency into read, regex, write components. Selective DCA disable lowers read portion.}
     \hfill
     \subfloat[System-wide I/O throughput (\textbf{F}astclick + \textbf{F}F\textbf{S}B-\textbf{H})\label{subfig:io_tp}]{\includegraphics[width=0.49\columnwidth]{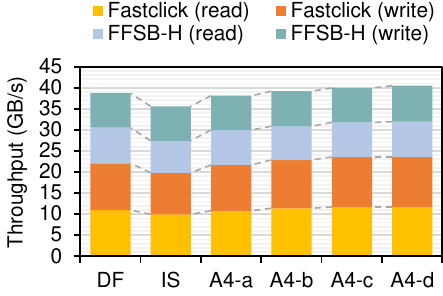}%
     }
     \Description{Stacked bars of system‑wide read/write throughput for Fastclick and FFSB‑H under each scheme. A4 maximizes combined throughput.}
     \hfill
     \subfloat[System-wide memory bandwidth consumption\label{subfig:mem_bw}]{\includegraphics[width=0.49\columnwidth]{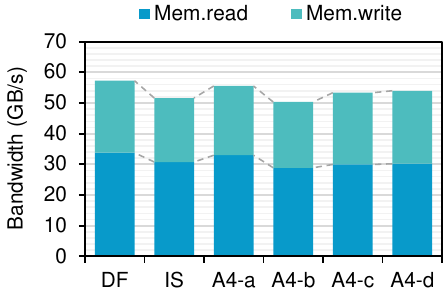}%
     }
     \Description{Bars of system memory read and write bandwidth for each scheme. A4‑d shows reduces memory bandwidth consumption.}
     \vspace{-5pt}
     \caption{I/O latency breakdown and system-wide performance metrics.}
     \label{fig:real_io_perf}
     \vspace{-7pt}
\end{figure*}
\begin{figure*}[!t]
     \captionsetup[subfigure]{justification=centering}
     \centering
     \subfloat[Partitioning thresholds\label{subfig:partition_param}]{\includegraphics[page=1, width=0.653\columnwidth]{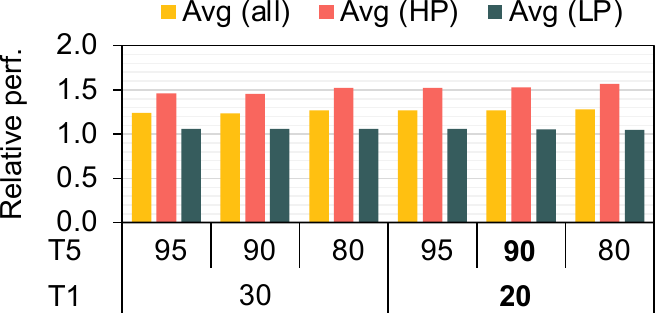}}
     \hfill
	\subfloat[Antagonist detection thresholds\label{subfig:DMA_leak_param}]{\includegraphics[page=3, width=0.615\columnwidth]{fig/sensitivity_study.pdf}}
     \hfill
	\subfloat[Timing parameter\label{subfig:timing_param}]{\includegraphics[page=2, width=0.635\columnwidth]{fig/sensitivity_study.pdf}}
     \Description{Performance vs.\ stable‑interval length with 1s revert. Curve approaches oracle as interval grows, 10s achieves 99.2\% of oracle performance.}
     \hfill
     \vspace{-9pt}
     \caption{Sensitivity study on \design parameters. Parameter values used in the main experiment are highlighted in boldface.}
     \label{fig:sensitivity_study}
     \vspace{-6pt}
\end{figure*}

\noindent \textbf{I/O latency and throughput.}
We use a modified Fastclick binary exclusively in \Fig{subfig:net_brkdwn} to measure network latency breakdown, as timestamping adds delay on the critical path.
We capture three parts: queueing in the Rx ring, accessing a packet pointer, and processing it.
\Fig{subfig:net_brkdwn} shows that \design-d reduces these three parts by 15\%, 20\%, and 23\% (vs. Default model).
In Fastclick, computation latency is reduced the most, indicating that I/O HPWs are well benefiting from DCA.
This matches the result depicted in ~\Fig{subfig:io_tp}, \ie, reduced latency translates into increased I/O throughput in network I/O device.
Since the I/O devices have reached the maximum available bandwidth, the gain in I/O throughput is less compared to that of I/O latency.
On the contrary, latency and throughput are largely unchanged in FFSB-H, supporting that it's insensitive to DCA or LLC caching.
The storage I/O breakdown in \Fig{subfig:stg_brkdwn} suggests that read latency is even larger when DCA is enabled (Default model) than when DCA is disabled (\design-c).

\noindent \textbf{Memory bandwidth.}
Pseudo LLC bypassing forces the dead cache lines to be written back to the memory, but the total number of write-backs remains largely unchanged (O5 in \Sec{sec:mitigation:dcaoff}).
This is depicted as a negligible difference in memory bandwidth in \Fig{subfig:mem_bw}, comparing \design-c and \design-d.
Disabling DCA for intensive storage I/O workload does increase the memory write bandwidth by 9\% from \design-b to \design-c.
However, overall \design-d reduces memory bandwidth by 6\% (vs. Default model).
Furthermore, despite the 5\% increase in the I/O throughput, \design-d reduces the memory read bandwidth by 11\%. (vs. Default model).
It is attributed to the improved LLC-caching efficiency for high locality data, and alleviating DMA leak. 
Note that, the low memory bandwidth of the Isolate model originates from reduced I/O throughput.

\noindent \textbf{Sensitivity study.}
We investigate the sensitivity of thresholds and parameters by categorizing them into three groups: (1) partitioning thresholds (T1 and T5), (2) I/O antagonist detection thresholds (T2---T4), and (3) timing parameters.
We perform a sensitivity study using an HPW-heavy configuration, highlighting main experimental parameters in bold.
\Fig{fig:sensitivity_study} illustrates relative performance of \design normalized to the Default model.

First, \Fig{subfig:partition_param} depicts the impact of T1 and T5, which establish the optimal partition of \design.
T1 defines the acceptable range of HPW performance.
As T1 decreases, the \design constrains the LP Zone, thereby enhancing the performance of HPWs while diminishing that of LPWs.
T5 is used to dynamically detect non-I/O antagonistic workloads.
At T5 values of 90\% and 95\%, \design identifies two non-I/O antagonists, whereas, at that of 80\%, it detects three, which leads to additional performance gain in HPWs.
However, this comes at the cost of a non-I/O HPW (\ie, {xalancbmk}), sacrificing its performance and failing to meet its QoS.
Therefore, despite an overall performance increase, a T5 threshold of 80\% does not appear optimal.

Second, in \Fig{subfig:DMA_leak_param}, T2, T3, and T4 jointly indicate that the certain I/O workload is not leveraging the advantage of DCA, causing severe DMA leak.
As the threshold values increase, FFSB-H is no longer identified as an antagonist, leading to suboptimal performance.
By gradually increasing each value, we pinpoint the threshold values at which FFSB-H ceases to be detected as an antagonist, and these critical thresholds are marked in red in \Fig{subfig:DMA_leak_param}.

Finally, in \Fig{subfig:timing_param}, we intentionally chose a stable phase in our experiment to investigate the overhead of periodic reverting.
\design periodically reverts to the initial partition, which might yield inferior performance if triggered too frequently.
As shown in \Fig{subfig:timing_param}, with a fixed revert interval of 1 second, as the stable interval increases, performance approaches the oracle policy where \design never reverts and stays in a stable state.
In particular, 10 seconds of stable interval achieves 99.2\% performance of the oracle policy.

\section{Related Work}
\label{sec:related_work}

{\noindent \textbf{Handling DMA bloat.}}
Existing solutions to address DMA bloat are (1) self-invalidating cache lines that invalidates consumed I/O buffers without writing them back to the LLC~\cite{alian2022idio,vemmou2022patching} and (2) replacing LLC-unfriendly cache lines earlier based on re-reference interval prediction~\cite{jaleel2010high,wu2011ship,qureshi2007adaptive,xie2009pipp,srinath2007feedback} or dead block prediction~\cite{khan2010using,jimenez2010dead,lai2001dead,liu2008cache}.
However, these approaches either entail a high overhead of detecting dead cache lines or require significant changes across the computing stack (\eg program codes, ISAs, and cache controllers).
As an alternative solution, we introduce pseudo LLC bypassing, which is readily applicable to commodity servers.

{\noindent \textbf{Mitigating latent contention.}}
ShRing~\cite{pismenny2023shring} proposes reducing the aggregate memory footprint of receive rings by sharing them among CPU cores, while DMA Cache~\cite{tang2010dma} suggests better isolation by separating on-chip storage for I/O data. 
Some studies~\cite{tootoonchian2018resq,yuan2021don} address the problem without such software or hardware modification by utilizing CAT-based cache way partitioning. 
Our work provides a holistic LLC orchestration framework that addresses not only these contentions but also two new I/O-driven contentions.
\section{Conclusion}
This work uncovers two new I/O-driven LLC contentions: between I/O and non-I/O workloads within inclusive ways and between storage-I/O and network-I/O workloads within DCA ways. 
Then, this work presents \design that orchestrates LLC sharing to holistically address these new I/O-driven LLC contentions when co-running diverse workloads, as well as previously identified ones.
It improves the performance of high-priority workloads by 51\% without notably compromising that of low-priority workloads.

\begin{acks}
This work was supported in part by the MSIT, Korea, under the Global Scholars Invitation Program (No. RS-2024-00456287) supervised by the IITP;
by the National Research Foundation of Korea (NRF) (No. RS-2025-00554650);
by the Technology Innovation Program (No. RS-2024-00420541, 2410000802) funded by the Ministry of Trade, Industry and Energy of Korea;
by the Yonsei University Research Fund of 2025-22-0104;
and by gift grants from Intel Corp.
Ipoom Jeong is the corresponding author.
\end{acks}
\appendix
\section{Artifact Appendix}

\subsection{Abstract}

In this section, we provide detailed instructions to reproduce the main experimental results presented in the paper.
The artifact includes three primary sets of experiments: (1) experiments revealing new I/O-driven LLC contentions (\Fig{fig:ioa_ma_interference}, \ref{fig:dir_lat_cont}, \ref{fig:leaky_dma_fio}, and \ref{fig:dpdk_fio_interference}); (2) experiments demonstrating the effectiveness of contention mitigation solutions (\Fig{fig:overlapping_ways}, and \ref{fig:device_aware_stategies}); and (3) experiments evaluating A4 with real-world workloads (\Fig{fig:perf_llc_hit_rate}, and \ref{fig:real_io_perf}).

\subsection{Artifact check-list (meta-information)}

{\small
\begin{itemize}
  \item {\bf Program:}
  
  - Microbenchmarks: DPDK-T, DPDK-NT, Flexible I/O Tester (FIO)~\cite{GitHubax76:online}, and X-Mem~\cite{GitHubmi87:online}
  
  - Real-world workloads: Fastclick~\cite{tbarbett20:online}, FFSB~\cite{FFSBPrim6:online}, Redis~\cite{Redis81:online, cooper2010benchmarking}, and SPEC CPU2017~\cite{spec2017}
  
  - Tools: Intel Performance Counter Monitor (PCM)~\cite{intelpcm30:online}, Intel CAT~\cite{GitHubin14:online}, and ddio-bench~\cite{farshin2020reexamining}
  \item {\bf Compilation: }
  GCC-10.0.5. Python 2.7 and Scons 2.3.0
  \item {\bf Binary: }
  DPDK-(N)T, X-Mem, and ddio-bench
  \item {\bf Run-time environment: }
  Ubuntu 20.04 LTS
  \item {\bf Hardware: }
  A server machine with Intel Xeon Gold 6140 CPU, and a client machine connected to 100Gbps DPDK-compatible NICs. Four M.2 SSDs and a M.2 NVMe RAID Controller.
  \item {\bf Metrics: } Throughput, average and p99 latency, LLC hit rate, memory bandwidth consumption, and IPC.
  \item {\bf Output: } Text files with raw data, and figures. 
  \item {\bf Experiments: } \Fig{fig:ioa_ma_interference}, \ref{fig:dir_lat_cont}, \ref{fig:leaky_dma_fio}, \ref{fig:dpdk_fio_interference}, \ref{fig:overlapping_ways}, \ref{fig:device_aware_stategies}, \ref{fig:perf_llc_hit_rate}, and \ref{fig:real_io_perf}
  \item {\bf How much disk space required (approximately)?: $\leq$ 20GB}
  \item {\bf How much time is needed to prepare workflow (approximately)?: 1h }
  \item {\bf How much time is needed to complete experiments (approximately)?: $\sim$ 10h }
  \item {\bf Publicly available?: \url{https://github.com/ece-fast-lab/ISCA-2025-A4}}
  \item {\bf Code licenses (if publicly available)?: MIT License}
  \item {\bf Archived (provide DOI)?: 10.5281/zenodo.15105163}
\end{itemize}
}

\subsection{Description}

\subsubsection{How to access}
\label{sec:access}
The artifact can be accessed on Github (https://github.com/ece-fast-lab/ISCA-2025-A4). Most of the benchmarks and tools are publicly available via the following links.

\begin{itemize}
    \item {\bf FIO:} https://github.com/axboe/fio
    \item {\bf X-Mem:} https://github.com/microsoft/X-Mem
    \item {\bf Fastclick:} https://github.com/tbarbette/fastclick
    \item {\bf FFSB:} https://github.com/FFSB-Prime/ffsb
    \item {\bf Redis:} https://github.com/redis/redis
    \item {\bf YCSB:} https://github.com/brianfrankcooper/YCSB
    \item {\bf PCM:} https://github.com/intel/pcm
    \item {\bf CAT:} https://github.com/intel/intel-cmt-cat
    \item {\bf DDIO-bench:} https://github.com/aliireza/ddio-bench 
\end{itemize}

\subsubsection{Hardware dependencies}
The experiment requires two machines equipped with DPDK-compatible 100 Gbps NICs connected to each other. The server machine should have I/O devices in the same socket with at least 18 cores. We conducted an evaluation on an Intel Xeon Gold 6140 CPU (Skylake) with Nvidia BlueField-2 and four Samsung 980 PRO 1TB M.2 SSDs with a RAID controller installed in the same node.
\subsubsection{Software dependencies}
Please set up the dependencies for benchmarks and tools according to each GitHub repository listed in \Sec{sec:access}. Python 3 is required to generate figures along with the matplotlib, numpy, and pandas libraries. SPEC CPU2017 suite is required to run real-world workload experiments.

\subsection{Installation}

First, clone the artifact repository.
\begin{lstlisting}
$ git clone https://github.com/ece-fast-lab/ISCA-2025-A4
$ cd ISCA-2025-A4
\end{lstlisting}

Second, install benchmarks and tools. Follow the detailed instructions in \texttt{app/README.md} and \texttt{tools/README.md}. Third, set the client machine as in \texttt{client/README.md}, and corresponding environment variables in \texttt{scripts/utils/env.sh} properly. 

Some benchmarks are modified, and we provided a forked repository of the official benchmark, which is also publicly available. Modifications of each benchmark are indicated in \texttt{app/README.md}. Also, we provide pre-compiled binaries of some benchmarks and tools along with the source codes, such as DPDK micromenchmarks (DPDK-(N)T), X-Mem with different core affinities, and ddio-bench.

\subsection{Experiment workflow}
\label{experiment}

Run the initial setup scripts every time the machines are rebooted.
\begin{lstlisting}
# in server machine
$ cd scripts/utils
$ source ./env.sh
$ sudo ./init.sh
# in client machine
$ sudo ./setup_basic.sh
\end{lstlisting}

We provide guidelines on \texttt{scripts/README.md} for running experiments. \texttt{run\_motivation.sh} runs figures 3 to 8, and \texttt{run\_evaluation.sh} runs figures 13 and 14.

\begin{lstlisting}
$ scripts/run_motivation.sh
$ scripts/run_evaluation.sh
\end{lstlisting}
\subsection{Evaluation and expected results}

Results generated from \Sec{experiment} demonstrate I/O-driven new contentions, suggest a mitigation technique, and evaluate \design on the real-world benchmarks. It improves the performance of HPWs significantly without notably compromising that of LPWs. Detailed observations and key takeaways for each experiment can be found under the corresponding folders in the repository. Since experiments are conducted on real hardware, results may vary due to several influencing factors. Specifically, X-Mem tends to exhibit inconsistent performance, which could result in trends not precisely matching those presented and necessitate repeated runs.






\bibliographystyle{ACM-Reference-Format}
\bibliography{reference}

\end{document}